\documentclass[aps,prd, preprint, nofootinbib, preprintnumbers]{revtex4-2}
\usepackage{graphics,color, hyperref}
\usepackage[latin9]{inputenc} 
\usepackage{amsmath}
\usepackage{amsfonts}
\usepackage{amssymb}
\usepackage{graphicx} 
\usepackage{slashed}
\global\long\def\tr{\mathrm{Tr}}
\global\long\def\bR{\mathbb{R}}
\global\long\def\dd{\mathrm{d}}
\global\long\def\not#1{\slashed{#1}}%
\usepackage[many]{tcolorbox}

\newcommand{\be}{\begin{equation}}
\newcommand{\ee}{\end{equation}}
\newcommand{\bea}{\begin{eqnarray}}
\newcommand{\eea}{\end{eqnarray}} 
\begin{document}
\title{ 
Existence of a supersymmetric extension of
the  gauged non-linear $\Sigma$-model and the transition from integrability to chaos
\\~}
\author{Fabrizio Canfora${}^{1,2}$}
\affiliation{${}^1$Facultad de Ingenieria, Arquitectura y Dise\~no, Universidad San Sebasti\'an, sede Valdivia, General Lagos 1163, Valdivia 5110693, Chile,
\\
${}^2$Centro de Estudios Cient\' ificos (CECs), Avenida Arturo Prat 514, Valdivia, Chile}
\email{fabrizio.canfora@uss.cl}
\author{Nicol\'as Grandi${}^{3,4}$}
\affiliation{${}^3$Departamento de F\'isica, Universidad Nacional de La Plata, \\ Casilla de Correos 67, La Plata, Argentina,
\\
${}^4$Instituto de F\'isica La Plata, Consejo Nacional de Investigaciones Cient\'ificas y T\'ecnicas, \\ Diagonal 113 esquina 63, La Plata, Argentina.}
\email{grandi@fisica.unlp.edu.ar}
\author{Marcelo Oyarzo${}^{5,6}$}
\affiliation{~${}^5$Departamento de F\'isica, Universidad de Concepci\'on,   Casilla, 160-C, Concepci\'on, Chile.
\\
${}^6$Department of Applied Science and Technology, Politecnico di Torino,
C.so Duca degli Abruzzi, 24, I-10129 Torino, Italy\vspace{.5cm}}  
\email{moyarzo2016@udec.cl}

\begin{abstract}  
The non-linear $\Sigma$-Model minimally coupled with Maxwell theory in $3+1$ dimensions possesses a topologically non-trivial sector characterized by \emph{lasagna}-like configurations. We demonstrate that, when a specific quantization condition is met, the associated second-order field equations admit a first-order Bogomol'nyi-Prasad-Sommerfield system. 
This discloses 
the existence of a $(1+1)$-dimensional supersymmetric extension
with $\mathcal{N}=2$ supercharges. 
We examine its 
imprint 
on the time-dependent regime, with particular emphasis on the transition from integrability to chaos.
\end{abstract}
\maketitle
\newpage 
\tableofcontents

\newpage 
\section{Introduction}
The gauged non-linear $\Sigma$-model is one of the most significant field theory models, with applications spanning from statistical mechanics and condensed matter systems to high-energy physics. Its versatility is evident in its ability to describe  phenomena such as quantum magnetism, the quantum Hall effect, and superfluidity, as well as  mesonic dynamics in the strong coupling limit \cite{NN[4],N[2],N[3], N[4]}.
In this manuscript, we analyze the gauged non-linear $\Sigma$-model in $3+1$ dimensions from the perspective of Quantum Chromodynamics (QCD), as it represents the low-energy limit of QCD minimally coupled to Maxwell electromagnetism. The modern approach to such a description is referred to as Chiral Perturbation Theory ($\chi$PT) \cite{N[43], N[44], N[45],N[46], N[47], N[48], N[49], N[50]}. 

Many interesting open problems appear when a large number of baryons are confined on a finite spatial volume. There are strong theoretical and phenomenological evidence  that, under such conditions, baryons melt and
inhomogeneous baryonic condensates appear \cite{pasta1, pasta2, pasta3, pasta4, q[12],q[8], canfora2, 56, 56b,58,58b, ACZ, CanTalSk1, canfora10, Fab1, gaugsk, Canfora:2018clt, lastEPJC, crystal1}  {(for pion and chiral condensates see e.g. \cite{ex4d6,ex4d4,ex4d5,ex4d1,ex4d2, ex4d3}}. These configurations share some features with the Larkin-Ovchinnikov-Fulde-Ferrell phase in superconductors \cite{LOFF1, LOFF2}. 
 
For these reasons, we focus on  the phase diagram of baryonic matter under extreme conditions, at finite baryon density and strong magnetic fields. From the viewpoint of QCD, such analysis is very difficult \cite{R0,  R11, R2}, since perturbative techniques are not effective at low energies, and lattice methods fail as it is still not clear how to avoid the sign problem \cite{sign1,sign2,sign3m}. 

In this paper, we demonstrate that, 
at finite baryon density and under suitable circumstances,
the model admits a supersymmetric extension, improving 
the integrability of the $\chi$PT equations. 
Of course, it is not a \emph{fundamental} supersymmetry  of nature, which would require that each elementary particle posses a supersymmetric partner of opposite
statistics \cite{Sohnius:1985qm}. It appears
when   the effective action is supplemented with a pair of Majorana spinors, which is possible only when a suitable quantization condition is satisfied.  

An interesting feature of the  
supersymmetric extension discussed here is that it cannot be implemented when the electromagnetic degrees of freedom are turned off. This is to be contrasted with the nice results in \cite{NittaSUSY} where the non-linear realization of supersymmetry  in the ungauged Skyrme model is discussed. Moreover, it requires a quatization condition on the area of the
magnetized baryonic layers. 
The supersymmetric theory possesses a central charge which is a non-linear function of the the $\Sigma$-model topological charge\footnote{ For the relevance of the barionic charge as topological invariant see \cite{Skyrme1,Skyrme2,Skyrme3,Skyrme4}, and \cite{gaugsol0,gaugsol,gaugsol1,gaugsol2,gaugsol3,gaugsol4,gaugsol5,gaugsol6,gaugsol7,gaugsol8} for analytic solution of Skyrme model and modification to it realization non-trivial topological charges.
  }  \cite{BPSlast1, BPSlast2} {\em i.e.} the baryonic charge. Moreover, the regular field configurations preserving supersymetry satisfy first order equations and have a non-vanishing value for such central charge. 

In order to disclose the results discussed above, we first derive a suitable Ansatz, which allows us to describe the gauge non-linear $\Sigma$-model configurations in terms of two scalar degrees of freedom, and we write the corresponding effective action. At a first glance, one could be rather pessimistic about the project, since it is known that the obvious Bogomol'nyi-Prasad-Sommerfield (BPS) bound in terms of the baryonic charge cannot be saturated. This is at odds with supersymmetry, which would require that
a BPS bound would be saturated in order to identify the central
charge of the superalgebra with the topological charge. In fact the
techniques introduced in \cite{BPSlast1, BPSlast2} allowed the derivation of BPS bounds in terms of topological charges which are non-trivial functions of the obvious charges. These results are the ingredients needed to disclose the supersymmetric extension of $\chi $PT at finite Baryons density.

One of the most interesting effects of the existence of a supersymmetric extension discussed here  is related to chaotic dynamics.  Starting from \cite{Chaos1} there has been a lot of recent activity in the analysis of chaos in field theory \cite{Chaos2}. For example, it was discovered that Yang-Mills theory possesses time-dependent configurations which are chaotic \cite{Chaos3, Chaos4}. The effect of non-trivial topological charges and fluxes on such chaotic behavior  has not been fully analyzed yet, some steps in this direction can be found in \cite{Canfora:2023syr}. The same question is very interesting also in the present case of the gauged non-linear $\Sigma$-model. In the forthcoming sections we will show that the emergence of SUSY is felt in the dynamics of the theory, as when the parameters of the theory are close to their supersymmetric values, the dynamics tends to be less chaotic.  
\newpage

\section{The gauged non-linear $\Sigma$-model}
\label{sec:model}
\subsection{Definitions and properties of the model}
The dynamical content of the gauged non-linear $\Sigma$-model is given by a $SU(2)$ valued scalar field $U$, which is coupled to an Abelian gauge connection $A$ {(see for instance \cite{Callan_Witten,sigma1,sigma2,sigma3} for the relevance of the model)}.  We can write a classical action for it, in the form
\begin{equation}
I\left[ U,A\right] =\frac{\kappa}{4}\,\text{tr}\!\int \Sigma \wedge \star \Sigma -%
\frac{1}{2}\int F\wedge \star F\ ,
\label{eq:Sigma.model}
\end{equation}
where the star $\star(\,\cdot\, )$ denotes the Hodge dual operation, and the one-form $\Sigma =U^{-1}DU$ is defined in terms of the gauge covariant derivative $D(\,\cdot\, ) =d(\,\cdot\,) +e\,A\left[ t_{3},\,\cdot\, \right]$. The electric charge $e$ can be re-scaled into $\kappa$, which is the only relevant coupling constant of the system. 
The role of the covariant coupling is to enhance the invariance of the neutral $\Sigma$-model under global $SU(2)$ transformations, to include gauge transformations in the direction of the $t_3$ generator. This is consistent with the existence of only one neutral pion.

The model has a conserved topological current $J$, 
defined by the 
combination 
\begin{equation}
J=\star \,\text{tr}\left[U^{-1}dU\wedge U^{-1}dU\wedge U^{-1}dU\right]-3\,\star d\left(
A\wedge\text{tr}\left[  t_{3}\left( U^{-1}dU+dUU^{-1}\right) \right]\right) \ ,
\label{eq:topological.charge}
\end{equation}
which is gauge invariant and
has a time component that, once integrated in a space-like hypersurface, gives the baryonic charge. 


We will be interested in topologically non-trivial configurations of the 
dynamics 
defined by action \eqref{eq:Sigma.model}. The corresponding equations of motion  take the form
\begin{eqnarray}
&&D\wedge \star \Sigma =0\ ,
\label{eq:equations.of.motion.1} 
\\
&&\star\, d\star F =
\frac{\kappa}{2}\,\text{tr}\left[U^{-1}\!\left[ t_{3},U\right]
 \Sigma \right] \ .\label{eq:equations.of.motion.2} 
\end{eqnarray}
We want to find solutions to these equations in flat space, with
non-vanishing topological charge \eqref{eq:topological.charge}. In order to do that, we will need to impose non-trivial boundary conditions, which we will ensure by adding to the action \eqref{eq:Sigma.model} a suitable boundary term, see below.

\newpage

\subsection{Ansatz and equations of motion}
Let us choose the spacetime coordinates as $(t,r,\vartheta,\varphi)$ in order to write the $3+1$ Minkowskian metric in the form
\begin{equation}
ds^{2}=\frac{1}{2\kappa}\left(-dt^{2}+dr^{2}\right)+\ell ^{2}\left( d\vartheta ^{2}+d\varphi ^{2}\right) \ .
\end{equation}
Here $0<\vartheta\leq 2\pi$ and $0<\varphi\leq 2\pi$ are Cartesian coordinates on a two-torus, and $\ell$ is the characteristic size of the box. {  The coordinates range are adapted to the Euler angles that we implement in the parameterization of the SU(2) valued field \cite{Euler1,Euler2,Euler3}}. On the other hand we choose $r$ on an arbitrary range. The $1/4\kappa$ scale is included for later convenience. 

For the $\Sigma$-model and gauge fields we propose the following Ansatz 
\begin{eqnarray}
&&U =e^{n\varphi \,t_{3}} e^{  v(r, t)
t_{2}} e^{n\vartheta  \,t_{3}} \ ,
\label{eq:anzatz.1}
\qquad\qquad\quad
A= \frac12 \left(n-\sqrt{2\kappa}\,\ell\,u(r, t) \right) \left( d\vartheta -d\varphi
\right) \ , 
\end{eqnarray}
where $(u,v)$ are two functions to be determined by the equations of motion, and $n\in \mathbb{Z}$. It can be shown that the energy and charge densities  resulting from the Ansatz \eqref{eq:anzatz.1} are constant on the  $(\vartheta, \varphi)$ plane. This allows us to interpret it as a baryonic layer, or in other words a nuclear \emph{lasagna} phase. { Observe that the Ansatz is 4-dimensional in the sense that the fundamental fields depends on the four spacetime coordinates, which allow them to have a non-trivial topological charge.}

The non-trivial dependence of the Ansatz on the three coordinates is chosen so that the topological current does not vanish. Indeed, from expression \eqref{eq:topological.charge}  we get 
%
\begin{equation} 
J=12\alpha\left(  \partial_t (u\cos^2\! v) \, dr + \partial_r (u\cos^2\! v) \, dt \right)\,.
\label{eq:topological.charge.Ansatz}
\end{equation}
Here we defined $\alpha=n/\sqrt{2\kappa}\ell$.
The corresponding topological charge would be non-vanishing as long as the time component integrates to a finite value in a space-like hypersurface, as
\begin{equation}
B=12 \alpha(2\pi\ell)^2
\left.(u\cos^2\!v)\right|_{\sf bdy}\,.
\label{eq:topological.charge.Ansatz}
\end{equation}
Where ``${\sf bdy}$'' refers to the boundaries of the $r$ interval. Regarding the electromagnetic strength, we have 
\begin{equation}
    F=-\frac n\alpha\,\partial_r u \, dr\wedge (d\vartheta-d\varphi)
    -\frac n \alpha\,\partial_t u \, dt\wedge (d\vartheta- d\varphi)\,,
\label{eq:electromagnetic.field.Ansatz}
\end{equation}
resulting in mutually perpendicular electric and magnetic fields lying on the $(\vartheta,\varphi)$ plane.

Replacing the Ansatz \eqref{eq:anzatz.1} 
  into the seven coupled field equations 
\eqref{eq:equations.of.motion.1}-\eqref{eq:equations.of.motion.2} we get a pair of non-linear coupled partial differential equations for our Ansatz variables $u$ and $v$, as 
\begin{eqnarray}
&&\square u- 2u\,\sin^2\!v=0\,,
\label{eq:equations.Ansatz.1}
\\
&&\square v+   \left(\alpha^2-u^2\right)\sin(2v)=0\,,
\label{eq:equations.Ansatz.2}
\end{eqnarray}
where $\square(\,\cdot\,)=-\partial_t^2(\,\cdot\,)+\partial_r^2(\,\cdot\,)$.

The equations above \eqref{eq:equations.Ansatz.1}-\eqref{eq:equations.Ansatz.2} can be derived from an effective $(1+1)$-dimensional action for the pair of scalar fields $u,v$ with the form
\begin{equation}
I_{\sf eff}[u,v]=-\int dt\,dr\,
\left(
\frac{1}{2}(\partial u)^2+\frac{1}{2}(\partial v)^2
+\left(u^{2}-\alpha^2\right)\sin ^{2}\!v 
\right)+I_{\sf bdy}[u,v]\,,
    \label{eq:effective.action}
\end{equation}
here $(\partial\,\cdot\,)^2=-(\partial_t\,\cdot\,)^2+(\partial_r\,\cdot\,)^2$, and the boundary term $I_{\sf bdy}[u,v]$ fixes the boundary conditions for the Ansatz functions $u$ and $v$.

\section{Supersymmetric extension}
\subsection{First order equations}
\label{sec:first.order.equations}
Interestingly, there is a set of first order equations that are equivalent to \eqref{eq:equations.Ansatz.1}-\eqref{eq:equations.Ansatz.2} for some particular choice of the box size $\ell$. They can be written as follows
\begin{eqnarray}
&&\partial_\mu u+\left(\sqrt 2 \,t_\mu+t_\mu^\perp\right) \cos v =0\,,
\label{eq:equations.first.order.1}
\\
&&\partial_\mu v-{\sqrt{2}}\,t_\mu\, u \sin v=0\,.
\label{eq:equations.first.order.2}
\end{eqnarray}
Here $t^\mu$ is a constant space-like unit vector $t_\mu t^\mu=1$ in the two dimensional Minkowskian plane $(t,r)$, and $t_\mu^\perp$ is any constant form perpendicular to it $t_\mu^\perp t^\mu=0$. 
Taking the divergence of these equations and replacing them in the result, we obtain
%
%
%
\begin{eqnarray}
&&\square u- 2  u \sin^2 v=0\,,
\label{eq:equations.Ansatz.fixed.1}
\\
&&\square v+\left(1-u^2
\,  
\right)\sin(2v)=0\,,
\label{eq:equations.Ansatz.fixed.2}
\end{eqnarray}
which coincide with \eqref{eq:equations.Ansatz.1}-\eqref{eq:equations.Ansatz.2} whenever $\alpha=1$, namely when the size of the box satisfies  
\begin{equation}
\ell =\frac{n}{\sqrt{2\kappa }} \, .  \label{QuantizationCondition}
\end{equation}
The above condition corresponds to a quantization of the area of the magnetized baryonic layers, which are the solitons of the theory.

Equations \eqref{eq:equations.first.order.1}-\eqref{eq:equations.first.order.2} imply that the functions $u,v$ are time independent. Indeed, since $t_\mu$ is a space-like unit vector we can boost to its rest frame $t_\mu=(0,-\epsilon)$ with $\epsilon=\pm 1$ is a sign. This results in a time-like perpendicular with the form $t_\mu^\perp=( b,0)$ with $b$ an arbitrary constant. In terms of the boosted coordinates, that for simplicity we denote with the same name $(t, r)$, we can then rewrite the above equations as
\begin{eqnarray}
&&\partial_r u {\,-\,}\sqrt 2 \,\epsilon  \cos v =0\,,
\qquad\qquad\quad\ \,
\partial_t u +
b\cos v =0\,,
\label{eq:equations.first.order.r.1}
\\
&&\partial_r v {\,+\,} { \sqrt{2}} \,\epsilon\, u \sin v=0\,,
\qquad\qquad\quad 
\partial_t v=0\,.
\label{eq:equations.first.order.r.2}
\end{eqnarray}
Consistency 
requieres that $b$ vanishes. Indeed, since $v$ is time independent, the equation for $\partial_r v$ implies that $u$ is also time independent. Then from the equation for $\partial_t u$ we obtain $b=0$. 

{ It is clear that the first order equations \eqref{eq:equations.first.order.r.1}-\eqref{eq:equations.first.order.r.2} are not completely equivalent to the second order equations \eqref{eq:equations.Ansatz.1}-\eqref{eq:equations.Ansatz.2}, but they capture a subset of time-indepentent solutions of the latter when the parameters satisfy the condition \eqref{QuantizationCondition}. In other words, the constitute a Bogomol\'{}nyi-Prasad-Sommerfield (BPS) system (see bellow). It is worth emphasizing that not all second order field equations possess a first order BPS system. Indeed, as it happens, for instance, in the Ginzburg-Landau theory for BCS superconductors, where there is first order BPS system only at critical coupling. For non-Abelian monopoles in the Georgi-Glashow model, one can find a first order BPS system which implies the second order field equations only when the Higgs potential vanishes. In general, one should expect that the requirement for a given system of second order field equations to possess a first order BPS set is very restrictive and, in most of the cases, it is not satisfied. 
On the other hand, when a first order BPS system exists, then not only it is much easier to solve, but also the corresponding solutions are topologically stable. We will show that our first order system enjoys some of these properties.}

When replaced into the topological charge \eqref{eq:topological.charge.Ansatz} the first order equations dictate  
\begin{equation}
B=
6\alpha(2\pi\ell)^2
\left.(u\,(\partial_ru)^2)\right|_{\sf bdy} \,.
\label{eq:topological.charge.first.order}
\end{equation}
On the other hand, this is a purely magnetic configuration, with
\begin{equation}
F=-n\, \partial_ru \,dr\wedge (d\vartheta-d\varphi)\,.
\label{eq:electromagnetic.field.first.order}
\end{equation}

\subsection{A solution by quadratures}

\begin{figure}
    \centering
    \includegraphics[width=0.95\textwidth]{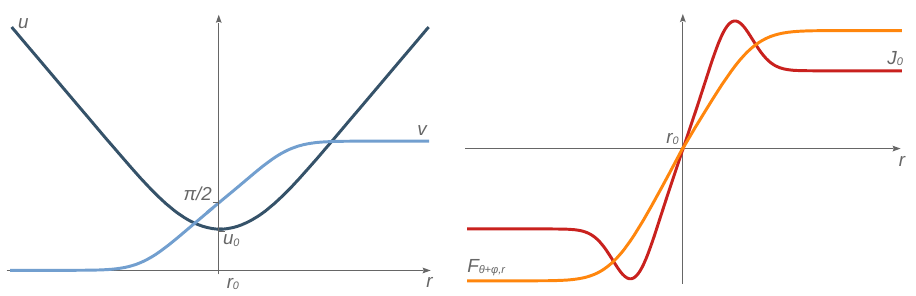} 
    \caption{ 
    Plots of the analytic solution \eqref{eq:sinv.quadratures}-\eqref{eq:u.quadratures} in the special point $\alpha=n/\sqrt{2\kappa}\ell=1$,  corresponding to $\epsilon_0=1$. \underline{Left}: profiles of the Ansatz functions $u$ and $v$. \underline{Right}: profiles of the observable quantities, the topological charge density $J_0$ and the magnetic field $F_{\vartheta+\varphi,r}$.}
    \label{fig:variables.and.observables}
\end{figure}

The first order equations \eqref{eq:equations.first.order.r.1}-\eqref{eq:equations.first.order.r.2} can be solved if we assume that {  $\partial_r v \neq 0$ and $u  \sin v \neq 0$, then the ratio between the equations \eqref{eq:equations.first.order.r.1} and \eqref{eq:equations.first.order.r.2} leads to a differential equation for $v$ as a function of $u$, that can be written in a simple manner by considering $v$ as a function of the variable $u^2$.} Indeed, writing $\partial_r v=v'\,2u\, \partial_r u$ in \eqref{eq:equations.first.order.r.2} and replacing $\partial_r u$ from \eqref{eq:equations.first.order.r.1}, we get
\begin{eqnarray} && 
2\,v'+ \tan v=0\,,
\label{eq:equation.udev}
\end{eqnarray}
where a prime $(\,\cdot\,)'$ means derivative with respect to $u^2$.
This equation for $v$ can be solved straightforwardly, obtaining 
\begin{equation}
    \sin v= \epsilon_0\, e^{-\frac12(u^2-u_0^2)} \, ,
    \label{eq:sinv.quadratures}
\end{equation}
where $u_0$ is an integration constant and $\epsilon_0=\pm1$ is a sign, and the solution is real whenever $u^2>u_0^2$. 
We can now write the remaining equation for $u$ by plugging this into \eqref{eq:equations.first.order.r.1}, which can then be solved by quadratures 
\begin{eqnarray}
r =r_0+ 
\int_{\epsilon_0u_0}^u \frac {\epsilon  \,du'}{\sqrt{2\left(1-e^{-(u'^2-u^2_0)}\right)}} \,.
\label{eq:u.quadratures}
\end{eqnarray}
Here $r_0$ is another constant of integration. We have two possible solutions according to the value of $\epsilon_0=\pm1$, one with $u>u_0$ and the other with $u<-u_0$. The value of $r$ is close to $r_0$ whenever the value of $u$ is close to $\epsilon_0u_0$. For large values of $u$, the integral in \eqref{eq:u.quadratures} scales linearly in $u$, implying $r\sim \epsilon\, u$. In other words, the two possible values for $\epsilon =\pm1$ do not correspond to different solutions, but instead describe the positive and negative $r$ semi-axes. Regarding the variable $v$,  the right hand side of \eqref{eq:sinv.quadratures} goes to $\epsilon_0=\pm1$ as $r$ approaches $r_0$, implying that $v=(\epsilon_0/2) \pi$. On the other hand, we have that $\sin v$ goes to zero as $|u|$ becomes large, from which we deduce that $v=0$ for $r\to\ -\infty$ and $v=\epsilon_0\,\pi$ for $r\to\infty$. Plots of the corresponding profiles are shown in Fig.\,\ref{fig:variables.and.observables} (left).
The resulting topological current 
and magnetic field
are shown in Fig.\,\ref{fig:variables.and.observables} (right).

Notice that the original Ansatz \eqref{eq:anzatz.1} is periodic on the $\vartheta$ and $\varphi$ directions. On the other hand, at $r\to\pm\infty$ our analytic solution  for the functions $v$ and $u$  satisfies Dirichlet and Neumann boundary conditions respectively. Moreover, the topological charge \eqref{eq:topological.charge.first.order} would vanish when the configuration is integrated along the whole $r$ axis, while it results in a finite value for any finite $r$ range. The convenience of these properties for phenomenological applications will be discussed on section \ref{sec:discussion}. 

\subsection{Bogomol\'{}nyi-Prasad-Sommerfield bound}
\label{sec:BPS.bound}
To explore the meaning of the obtained result, let us write the effective action \eqref{eq:effective.action} for the particular choice of $\alpha=n/\sqrt{2\kappa}\ell=1$ that allows for the first order form of the equations. We get
\begin{equation}
\!\!\!
I_{\sf eff}[u,v]=-  \int \! dr\,dt
\left(
\frac{1}{2}(\partial u)^2+\frac 1 2(\partial v)^2
+\left(u^{2}-1\right) \sin ^{2}\!v 
\right)- \int dt\left.\left(r+2 \,(u\,\cos v)\right)\right|_{\sf bdy} ,
    \label{eq:effective.action.BPS}
\end{equation}
where 
we included the correct boundary term to reproduce the boundary conditions satisfied by the quadrature solution \eqref{eq:sinv.quadratures}-\eqref{eq:u.quadratures}.  
We can now try to write the integrand as the sum of the squares of the radial equations in \eqref{eq:equations.first.order.r.1}-\eqref{eq:equations.first.order.r.2}, resulting in the form
\begin{eqnarray}
I_{\sf eff}[u,v]&=&-\frac12\int  \!dr\,dt
\left(
\left(
\partial_\mu u+\sqrt 2 \,t_\mu \cos v
\right)^2
+
\left(
\partial_\mu v
-\sqrt 2\,t_\mu u\sin v
\right)^2
\right)
\,.
    \label{eq:effective.action.BPS.squares}
\end{eqnarray} 
Here $t_\mu=(0,1)$ is a unit spacelike vector as in section \ref{sec:first.order.equations}.
We see that there is no total derivative corresponding to the barionic charge \eqref{eq:topological.charge.first.order}. 

Specializing the action \eqref{eq:effective.action.BPS.squares} to a static situation, we can write a Bogomol\'{}nyi-Prasad-Sommerfield bound for the total energy in the form $E=-I_{\sf eff}[u,v]\geq 0$. This bound is saturated whenever the first order equations \eqref{eq:equations.first.order.1}-\eqref{eq:equations.first.order.2} are satisfied.

The presence of a  Bogomol\'{}nyi-Prasad-Sommerfield bound strongly suggest that the system can be complemented with a set of fermionic degrees of freedom in such a way to construct a supersymmetric theory. We explore this issue in the next subsection.

\subsection{Supersymmetry}
The simplest strategy to construct a supersymmetric theory is to complement each of our bosonic fields $u$ and $v$ with a fermionic counterpart, which encode in the two-component Majorana spinors $\psi $ and $\chi $. Notice that we are augmenting the field content of theory with fermions that {\em a priori} are non-physical and are only needed to discluse the supersymmetry transformations.

Supersymmetric transformations are defined in terms of a fermionic parameter $\epsilon$, which is also a two-component Majorana spinor. The variation of the bosonic fields can then be written as
\begin{equation}
    \delta v=\bar\epsilon \psi\,,
    \qquad\qquad\quad
    \delta u=\bar\epsilon \chi\,.
    \label{eq:bosonic.susy}
\end{equation}
%
Regarding the variation of the fermionic fields $\psi$ and $\chi$, we would like to recover our  Bogomol\'{}nyi-Prasad-Sommerfield first order equations \eqref{eq:equations.first.order.1}-\eqref{eq:equations.first.order.2} from the condition of a purely bosonic configuration which preserves half of the supersymmetries. We then write
\begin{equation}
\delta\psi 
=\left(\not{\partial}v- \sqrt{2}\,u\sin v\right)\epsilon\,,
\qquad\qquad\quad
\delta\chi  
=\left(\not{\partial}u+ \sqrt{2}\cos v\right)\epsilon\,.
\label{eq:fermionic.susy}
\end{equation}
With this, for a purely space dependent $\partial_t u=\partial_t v=0$ and bosonic $\psi=\chi=0$ configuration, the condition of vanishing  supersymmetry variations for a non trivial $\epsilon$ spinor satisfying $\gamma^{1}\epsilon=\pm\epsilon$ reduces to the first order equations \eqref{eq:equations.first.order.r.1}-\eqref{eq:equations.first.order.r.2}.

In order to construct a supersymmetric theory under the above defined supersymmetry transformations \eqref{eq:bosonic.susy}-\eqref{eq:fermionic.susy}, we need to supplement the purely bosonic action $I_{\sf eff}[u,v]$ in  \eqref{eq:effective.action.BPS} with a fermionic counterpart $I_{\sf fer}[\psi,\chi]$ determinaning the dynamics of the two spinor fields $\psi$ and $\chi$, in such a way that the total action  
\begin{equation}
    I[u,v,\psi,\chi]=I_{\sf eff}[u,v]+I_{\sf fer}[\psi,\chi]\,,\label{eq:action.total}
\end{equation} 
is invariant up to a boundary term. This is satisfied by the expression
\begin{align}
    I_{\sf{fer}}  =-\frac{1}{2}\int\,dt\, dr
    &\left( 
    \overline{\psi}\not{\partial}\psi
    +\overline{\chi}\not{\partial}\chi
    +\sqrt{2}\,u\cos v\,\overline{\psi}\psi
    +2
    \sqrt{2}\sin v\,\overline{\psi}\chi
    \right)\,.
\label{eq:action.fermionic}
\end{align}

From action \eqref{eq:action.total} we can derive the Noether (super)charges associated to supersymmetry invariance, and compute the corresponding algebra. It turns out that the central charge coincides with the boundary term in \eqref{eq:effective.action.BPS.squares}, as expected for  a Bogomol'nyi-Prasad-Sommerfeld bound in agreement with \cite{BPSlast1} . The details of the calculation are given in Appendix \ref{sec:appendix.susy}.

{ 
It is worth emphasizing that our model is not inherently supersymmetric, but admits a supersymmetric extension within the effective two-dimensional sector describing a specific set of configurations, where relation \eqref{QuantizationCondition} holds. 
This modification of the field content is a usual tool in the literature on Bogomol\'{}nyi-Prasad-Sommerfield solitons, to show that such solutions preserve some of the supersymmetries of the extended model.}

\section{Chaotic behavior}

\subsection{Mechanical analogs}
The system defined by action \eqref{eq:effective.action} would behave as a mechanical systems whenever we choose the dependence of the variables $u,v$ to be purely temporal.
%
The corresponding mechanical action reads
\begin{equation}
I_{\sf mech}[u,v]=\int dt\,dr\,
\left(
\frac{1}{2}(\partial_t u)^2+\frac{1}{2}(\partial_t v)^2
-\left(u^{2}-\alpha^2\right)\sin ^{2}\!v 
\right)\,.
    \label{eq:effective.action.chaos}
\end{equation}
%
%
It is convenient to switch into a Hamiltonian description, obtaining 
\begin{equation}
    H=\frac12p_u^2+\frac 12 p_v^2
+
\left(u^{2}-\alpha^2\right)\sin ^{2}\!v\,. \label{Hamiltonian_mechanical_system}
\end{equation}
%
%
The corresponding energy is conserved, resulting in the motion being confined to a three-dimensional hypersurface $H=E$ in phase space. The potential, given by the last term in equation \eqref{Hamiltonian_mechanical_system}, exhibits multiple potential wells. The minima of these wells are located at points $(u,v)=(0, (k+1/2)\pi)$, where $k\in\mathbb{Z}$, and the potential takes the value $-\alpha^2$ there. Potential barriers separate the wells, extending along the $u$ direction at values of $v=k\pi$, where $k\in\mathbb{Z}$. On these barriers, the potential vanishes. Consequently, for negative energies the motion will remain confined within a single potential well. Furthermore, when we specify the initial conditions for the variables $(p_u,p_v, v)=({p_u}_{\sf in}, {p_v}_{\sf in}, {v}_{\sf in})$, the condition of a real $u_{\sf in}$ sets a lower bound on the energy. This results in
\begin{equation}
0>
E
>\frac12{p_u}_{\sf in}^2+\frac 12 {p_v}_{\sf in}^2-\alpha^2\sin ^{2}\!v_{\sf in}
\,.
\end{equation} 
This restriction allows us to fix the value of the energy once the initial conditions and the value of $\alpha$ are known, in such a way that the motion is bounded. With this at hand we can plot the corresponding Poincar\'e sections to determine whether the system shows chaotic behavior.  

\begin{figure}
    \centering 
 \includegraphics[width=0.99\textwidth]{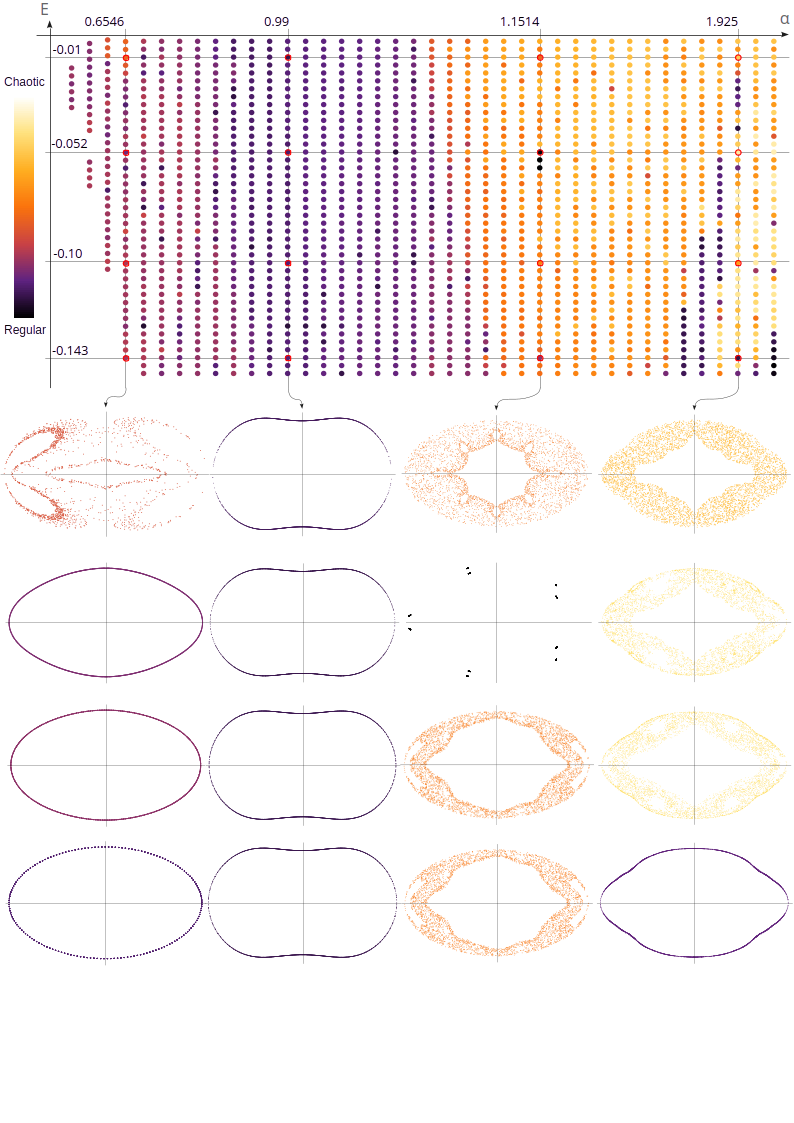} \vspace{-3.4cm} \caption{Phase diagram and Poincar\'e sections of the time dependent system, with initial conditions are $p_v(0) = 0.636$, $p_u(0) = 0.127$ and $v(0)=1.14$, while $u(0)$ is determined by the energy. We see that the Poincar\'e sections for parameters close to the Bogomol\'{}nyi-Prasad-Sommerfield value $\alpha=1$ are remarkably stable, while away from this point the system becomes chaotic at high enough energies.}
    \label{fig:poincaresections_set}
\end{figure}

In Fig. \ref{fig:poincaresections_set}, the chaotic behavior is quantified by the mean standard deviation of the Poincar\'e section of the corresponding configuration. The upper plot is a phase diagram which reveals two distinct regions, one of them highly chaotic {\em i.e.} showing large values of the standard deviation (depicted in yellow) and another with more regular behavior, corresponding to small standard deviation, (depicted in violet). We interpret the boundary between these two regions as transition to chaos.

Although the Bogomol\'{}nyi-Prasad-Sommerfield condition $\alpha=n/\sqrt{2\kappa}\,\ell=1$ does not allow to find real analytic solutions for the purely time dependent case, our numerical results provide evidence that this value for $\alpha$ retains some significance. Indeed,  the chaotic behavior of the system is suppressed as $\alpha$ gets close to its supersymmetric value. The corresponding Poincar\'e sections are plotted in the second column on the lower part of the Figure, and show that the system behaves quite regularly around this region of parameters.
\vspace{-.3cm}
\section{Discussion}
\label{sec:discussion}
\vspace{-.3cm}
Let us summarize the main results of the present work. First, we have shown that the non-linear $\Sigma$-model in $(3+1)$ dimensions, when confined inside a box whose size satisfies a suitable quantization condition, possesses a $(1+1)$-dimensional sector that admits a supersymmetric extension. Second, we obtain numerical evidence to support the claim that, for box sizes which are close to those quantized values, the time dependent solutions are non-chaotic.  To the best of authors' knowledge, this is the first result of this type in a theory  related with the low energy limit of QCD. 

We analyzed the gauged non-linear $\Sigma$-model under and Ansatz representing a crystal of baryonic layers or ``nuclear \emph{lasagna} phase''. The configurations of this sector are characterized by a single parameter $\alpha$ which depends on the $\Sigma$-model coupling $\kappa$, the length scale $\ell$ representing the periodicity in the longitudinal directions, and an integer number $n$ corresponding to a non-vanishing topological invariant. Mutually perpendicular electric and magnetic fields of the same strength, extend along the direction of the layers. 
When the parameter $\alpha$ takes the value $\alpha=1$, the equations of motion can be obtained from a first order system, which possesses analytic solutions.
 
The resulting analytic expressions for the Ansatz functions $u$ and $v$ satisfy Neumann ($u'=$ constant) and Dirichlet ($v=$ constant) boundary conditions, respectively, at large distances. In terms of physical observables, these conditions correspond to fixing the baryonic density and the magnetic flux. Notably, these conditions are natural from a phenomenological perspective. In actual neutron stars, various non-homogeneous baryonic condensates appear at slightly different densities, layered one on top of the other. At low densities, the formation of nuclear \emph{gnocchi} is expected. As we move closer to the star's center, the density increases, and the \emph{spaghetti} phase emerges. At even higher densities, nuclear \emph{lasagna} layers appear. It is believed that at densities exceeding those of the nuclear layers, the \emph{lasagna} phase transitions into a homogeneous distribution.
 
Our analytic solution yields a total baryonic charge of zero across the entire space, as it interpolates between a region of positive, constant baryonic density (fully populated by baryons) and another region of negative baryonic charge (dominated by anti-baryons). In actual phenomenological applications, the presence of a baryonic chemical potential compels the system to consist predominantly of baryons, with a negligible quantity of anti-baryons. To apply our solutions in such contexts, the boundaries of the region of interest should be positioned asymmetrically with respect to the point where the baryonic charge density vanishes. This asymmetry must be implemented under the constraint that the integral of the baryonic density returns an integer value. Notice that such number is non-linearly related to the parameter $n$ defining the solution.  

The existence of a special point in parameter space, where the equations can be derived from a first-order system, suggests the presence of a Bogomol'nyi-Prasad-Sommerfield (BPS) bound on the configuration's energy. We successfully formulated this bound using the effective action that provides the reduced equations for our Ansatz functions. Interestingly, in most systems exhibiting Bogomol\'{}nyi-Prasad-Sommerfield solitons, these solutions emerge at specific points in the coupling space where there is a balance between static repulsive (gauge) forces and attractive (gravitational or Higgs) ones. In contrast, the Bogomol\'{}nyi-Prasad-Sommerfield solitons in the present case are those whose size $\ell$ is commensurable in Fermi units $\ell=n/\sqrt{2\kappa}$.

Fist order Bogomol\'{}nyi-Prasad-Sommerfield equations are frequently encountered in supersymmetric systems as conditions for preserving some supersymmetry in the classical background. In our case, we demonstrated that the  $(1+1)$-dimensional system admits a ${\cal N}=2$ supersymmetric extension at finite baryon density. In other words, the configurations described by our Ansatz are organized into a supersymmetric multiplet.
It is worth emphasizing that the presence of the electromagnetic degree of freedom $u$ is needed as, without it, the supersymmetric extension would not be possible. This is to be contrasted with the results in \cite{NittaSUSY} where the non-linear realization of supersymmetry in the ungauged Skyrme model has been analysed. 

We explored the consequences of the existence of a supersymmetric extension on purely time-dependent configurations. Specifically, we examined how the transition into chaos is influenced by the proximity of the parameters to the quantization condition. We found that the closer the parameters are to this condition, the less chaotic the corresponding dynamics becomes. Thus, the existence of a supersymmetric extension leaves an imprint on the system's time evolution.

\acknowledgements
This work is partially supported by Fondecyt grants 1240048, and by CONICET grant PIP-2023-11220220100262CO and UNLP grant 2022-11/X931.
The Centro de Estudios Cient\'\i ficos (CECs) is funded by the Chilean Government through the Centers of Excellence Base Financing Program of Conicyt. MO is partially funded by Beca ANID de Doctorado grant 21222264.

\newpage

\appendix

\section{Embedding in a supersymmetric model}
\label{sec:appendix.susy}
\subsection{Definitions and spinorial identities }

%
We define the higher rank $\gamma$-matrices as $\gamma^{\mu_{1}\mu_{2}\dots\mu_{r}}\equiv\gamma^{[\mu_{1}}\gamma^{\mu_{2}}\dots\gamma^{\mu_{r}]}$.
The charge conjugation matrix $C$ is defined such that for any $\Gamma^{(r)}\in\{1,\gamma^{\mu},\gamma^{\mu_{1}\mu_{2}},\dots,\gamma^{\mu_{1}\dots\mu_{r}}\}$
\begin{align}
(C\Gamma^{(r)})^{T} & =-t_{r}C\Gamma^{(r)}\,,
\end{align}
where $t_{r}=\pm1$ depending on the rank of the $\gamma$-matrix.
All the coefficients are fixed in terms of $t_{0}$ and $t_{1}$, as $t_{2}=-t_{0}$, $t_{3}=-t_{1}$ and $t_{r+4}=t_{r}$.

The Majorana condition on a spinor $\lambda$ can then be imposed by requiring that the Majorana conjugated spinor, defined as $\overline{\lambda}_{c}=\lambda^{T}C$, coincides with the Dirac conjugated one $\overline{\lambda}=i\lambda^{\dagger}\gamma^{0}$. For arbitrary Majorana spinors $\psi$ and $\chi$ the following identity,
usually called \emph{Majorana flip}, holds
\begin{align}
\overline{\psi}\gamma^{\mu_{1}\dots\mu_{r}}\chi & =t_{r}\overline{\chi}\gamma^{\mu_{1}\dots\mu_{r}}\psi\,.
\end{align}
For even spacetime dimension $D=2m$ with $m\in\mathbb{N}$ and for $\Gamma^{A}\in\{1,\gamma^{\mu},\gamma^{\mu_{1}\mu_{2}},\gamma^{\mu_{1}\dots\mu_{r}}\}$
such that $\tr(\Gamma^{A}\Gamma_{B})=2^{m}\delta_{B}^{A}$, the \emph{Fierz rearrangement} identity can be written as
\begin{align}
(\overline{\lambda}_{1}M\lambda_{2})(\overline{\lambda}_{3}M^{\prime}\lambda_{4}) & =-\frac{1}{2}\sum_{A}(\overline{\lambda}_{1}M^{\prime}\Gamma_{A}M\lambda_{4})(\overline{\lambda}_{3}\Gamma^{A}\lambda_{2})\,.
\end{align}

\paragraph{Two-dimensional case: }
In the particular case $D=2$ we consider the Majorana basis for the $\gamma$-matrices, defined according to
\begin{align}
\gamma^{0}=i\sigma_{2}=\left(\begin{array}{cc}
0 & 1\\
-1 & 0
\end{array}\right)\,,\qquad\qquad\gamma^{1}=\sigma_{1}=\left(\begin{array}{cc}
0 & 1\\
1 & 0
\end{array}\right)\,,
\end{align}
Then the charge conjugation matrix reads
\begin{align}
C=-\sigma_{2} & \quad\implies\quad t_{0}=1\,,\quad\quad t_{1}=-1\,,
\end{align}
For this choice, a Majorana spinor has real components. We write down some useful identites for Majorana spinors $\lambda,\psi$ and a scalar field $v$ as follows
\begin{align}
&\overline{\lambda}\psi =\overline{\psi}\lambda\,,
&\quad&
\overline{\lambda}\gamma^{\mu}\psi=-\overline{\psi}\gamma^{\mu}\lambda\,,
\quad&
\\
&\overline{\lambda}\gamma^{\mu}\gamma^{\nu}\psi =\overline{\psi}\gamma^{\nu}\gamma^{\mu}\lambda\,,
&\quad&
\overline{\lambda}\gamma^{\mu\nu}\psi=-\overline{\psi}\gamma^{\mu\nu}\lambda\,,
&\quad&
\not{\partial}v\not{\partial}\psi=\partial^{\mu}v\partial_{\mu}\psi+\partial_{\mu}(\not{\partial}v\gamma^{\mu}\psi-\partial^{\mu}v\psi)
&~\nonumber
\end{align}
In two spacetime dimensions it is possible to match the off-shell degrees of freedom by considering 2 real scalars and 1 Majorana spinor. 

Let us define the chiral components $\psi_{+},\psi_{-}$ in terms
of the chiral $\gamma$-matrix $\gamma_{*}$
\begin{align}
\gamma_{*} & =\gamma^{0}\gamma^{1}=\left(\begin{array}{cc}
1 & 0\\
0 & -1
\end{array}\right)\,,\quad\text{such that}\quad(\gamma_{*})^{2}=1\,,\quad\{\gamma_{*},\gamma_{\mu}\}=0\,.
\end{align}
The definition of the chiral components is $\psi_{\pm}=\frac{1\pm \gamma_*}2\psi$. 

\paragraph{Graded Poisson algebra:}

The graded Poisson bracket $\{\cdot,\cdot\}_{PB}: \mathcal{F_{P}}\times\mathcal{F_{P}}\to\mathcal{F_{P}}$ is defined on functions $\mathcal{F_{P}}$ from a phase space $\mathcal{P}$ with Grassmann odd and Grassman even fields,  and must satisfy the following conditions
\begin{align}
&\{A,B\}_{PB}  =-(-1)^{g_{A}g_{B}}\{B,A\}_{PB}&\quad&
g_{{\{A,B\}}_{PB}}  =g_{A}+g_{B}\mod2\,,\nonumber \\
&\{AB,C\}_{PB}  =A\{B,C\}_{PB}+(-1)^{g_{B}g_{C}}\{A,C\}_{PB}B\,,
&\quad&~\label{leibniz rule condition}
\end{align}
where $g_{A}$ is equal to 0 for $A$ begin a Grassmann even field and to 1 for $A$ being a Grassmann odd one. To construct explicitly this bracket in a concrete setup, let us consider Minkowksi space $\bR^{1,d}$
with coordinates $(t,x^{i})$ and metric $\dd s^{2}=-\dd t^{2}+\dd\overrightarrow{x}^{2}$.
Let $(u^{a},p_{a})$ being Grassmann even bosonic fields and $(\psi_{\alpha},\pi^{\alpha})$ Grassmann odd fermionic fields. The definition of the equal time Poisson bracket is
\small
\begin{align}
\{A,B\}_{PB} & \equiv\int\dd^{d}x\ A\left(\frac{\overleftarrow{\delta}}{\delta u^{a}(x)}\frac{\overrightarrow{\delta}}{\delta p_{a}(x)}-\frac{\overleftarrow{\delta}}{\delta p_{a}(x)}\frac{\overrightarrow{\delta}}{\delta u^{a}(x)}\right.+
\left.\frac{\overleftarrow{\delta}}{\delta\psi_{\alpha}(x)}\frac{\overrightarrow{\delta}}{\delta\pi^{\alpha}(x)}+\frac{\overleftarrow{\delta}}{\delta\pi^{\alpha}(x)}\frac{\overrightarrow{\delta}}{\delta\psi_{\alpha}(x)}\right)B\,,
\end{align}
\normalsize
where sum on the repeated indices is understood and the left and right arrows on the derivatives denotes on which side it is acting, as needed to satisfy the Leibniz rule in the second line of (\ref{leibniz rule condition}). 

\subsection{Supersymmetric theory in two spacetime dimensions}
\paragraph{Free theory construction: }
Let $u,F$ be two real scalar fields and $\psi$ a Majorana spinor. The $1+1$ dimensional free action defined as
\begin{align}
I_{\sf{free}} & =\int\dd^{2}x\left(-\frac{1}{2}(\partial v)^{2}+\frac{1}{2}F^{2}-\frac{1}{2}\overline{\psi}\not{\partial}\psi\right)\,,
\end{align}
is invariant up a boundary term under the supersymmetry transformations 
\begin{align}
&\delta v  =\overline{\epsilon}\psi\,,
&\qquad&
\delta\psi  =\not{\partial}v\epsilon+F\epsilon\,,
\nonumber \\
&\delta F =\overline{\epsilon}\not{\partial}\psi\,,
&\qquad&
\delta\overline{\psi}  =-\overline{\epsilon}\not{\partial}v+F\overline{\epsilon}\,,
\end{align}
where $\epsilon$ is a Majorana spinor. The boundary term reads
\begin{align}
\delta I_{\sf{free}} & =\int\dd^{2}x
\ 
\overline{\epsilon}
\,\partial_{\mu}\!\left(\frac{1}{2}\not{\partial}v\gamma^{\mu}\psi-\partial^{\mu}v\psi+\frac{1}{2}\gamma^{\mu}\psi F\right)\,.
\end{align}
The proof of this requires the $D=2$ Majorana flip identity $\overline{\psi}\gamma^{\mu}\epsilon=-\overline{\epsilon}\gamma^{\mu}\psi$.

Considering the theory for $(v,F,\psi)$ presented above, the commutator of two successive supersymmetry variations generated by the two spinor parameters $\epsilon_1,\epsilon_2$ acting on the different fields, reads
\begin{align}
&[\delta_{2},\delta_{1}]v  =2\overline{\epsilon}_{1}\gamma^{\mu}\epsilon_{2}\,\partial_{\mu}v\,,
\qquad
&[\delta_{2},\delta_{1}]F  =2\overline{\epsilon}_{1}\gamma^{\mu}\epsilon_{2}\,\partial_{\mu}F\,,
\qquad
&[\delta_{2},\delta_{1}]\psi  =2\overline{\epsilon}_{1}\gamma^{\mu}\epsilon_{2}\,\partial_{\mu}\psi\,,
\end{align}
implying that they close on a spacetime translation with parameter $2\bar \epsilon_1\gamma^\mu\epsilon_2$.
The proof of these brackets requires some identities that hold in $D=2$. We sketch here some of the most relevant steps
\begin{align}
[\delta_{1},\delta_{2}]\psi & =\gamma^{\mu}\epsilon_{2}\overline{\epsilon}_{1}\partial_{\mu}\psi+\epsilon_{2}\overline{\epsilon}_{1}\not{\partial}\psi-(1\leftrightarrow2)\,,\\
 & =-\frac{1}{2}\sum_{A}\left(\Gamma_{A}\gamma^{\mu}+\gamma^{\mu}\Gamma_{A}\right)\partial_{\mu}\psi\,\overline{\epsilon}_{1}\Gamma^{A}\epsilon_{2}-(1\leftrightarrow2)\,,\nonumber \\
 & =-(\gamma_{\nu}\gamma^{\mu}+\gamma^{\mu}\gamma_{\nu})\,\partial_{\mu}\psi\,\overline{\epsilon}_{1}\gamma^{\nu}\epsilon_{2}-(\gamma_{\rho\sigma}\gamma^{\mu}+\gamma^{\mu}\gamma_{\rho\sigma})\,\partial_{\mu}\psi\,\overline{\epsilon}_{1}\gamma^{\rho\sigma}\epsilon_{2}\,,\nonumber \\
 & =-2\overline{\epsilon}_{1}\gamma^{\mu}\epsilon_{2}\,\partial_{\mu}\psi\,.\nonumber 
\end{align}
Here in the second equality we used a Fierz rearrangement where sum runs over $\Gamma_{A}\in\{1,\gamma_{\mu},\gamma_{\mu\nu}\}$. In the third equality we used the fact that $t_{0}=1$ and then, due to the anti-symmetry,
the term with the identity cancels, leaving only two non-trivial terms. The second term is zero due to the fact
that $\{\gamma^{\mu},\gamma_{\mu\rho}\}  =\gamma^{\mu}{}_{\nu\rho}=0$, which is true in two spacetime dimensions since $\gamma^{\mu}{}_{\nu\rho}$ is completely
anti-symmetric.
\paragraph{Adding integractions: }
So far we have a free theory for the fields $(v,F,\psi)$
where $v$ and $F$ are real scalar fields and $\psi$ is a Majorana spinor. We can now introduce a second set of fields $(u,G,\chi)$ in order to construct an interacting theory involving $(v,F,\psi)$ and $(u,G,\chi)$
that is supersymmetric. To be concrete, we consider the action $I =I_{\sf{free}}+I_{\sf{int}}$ with
\begin{align}
&I_{\sf{free}}  = \int\dd^{2}x\left(-\frac{1}{2}(\partial v)^{2}-\frac{1}{2}\overline{\psi}\not{\partial}\psi+\frac{1}{2}F^{2}-\frac{1}{2}(\partial u)^{2}-\frac{1}{2}\overline{\chi}\not{\partial}\chi+\frac{1}{2}G^{2}\right)\,,\label{free}
\\
&I_{\sf{int}} = \int\dd^{2}x\left(M(u,v)+N(u,v)F+L(u,v)G+\overline{\psi}\psi X(u,v)+\overline{\chi}\chi Y(u,v)+\overline{\psi}\chi Z(u,v)\right)\,,
\end{align}
The free part of the action $I_{\mathrm{free}}$ is then invariant under the combined supersymmetry transformations transformations of the multiplets $(u,F,\psi)$ and $(v,G,\chi)$, as
\begin{align}
& \delta v   =\overline{\epsilon}\psi\,,
&\qquad&
\delta\psi   =(\not{\partial}v+F)\epsilon\,,
\nonumber\\
& \delta F =\overline{\epsilon}\not{\partial}\psi
&\qquad& \delta\overline{\psi}  =\overline{\epsilon}(-\not{\partial}v+F)\,.\nonumber\\
&\delta u  =\overline{\epsilon}\chi\,, 
&\qquad&\delta\chi  =(\not{\partial}u+G)\epsilon\,,  
\nonumber\\
&\delta G  =\overline{\epsilon}\not{\partial}\chi &\qquad&\delta\overline{\chi}  =\overline{\epsilon}(-\not{\partial}u+G)\,, \label{SUSY var u and v}
\end{align}
We want to find the coupling interaction functions $M,N,L,X,Y,Z$ such that the interaction term in the action is also invariant under the same transformations.
To that end, it will be useful to redefine the terms on the interaction action $I_{\mathrm{int}}$ according to the power of the fermionic fields, as
\begin{align}
I_{\sf{int}}
 & =\int\dd^{2}x\left(\mathcal{L}_{M,N,L}+\mathcal{L}_{\overline{\psi}\psi}+\mathcal{L}_{\overline{\chi}\chi}+\mathcal{L}_{\overline{\psi}\chi}\right)\,.
\end{align}
And then analyze independently the variation of each term. We get%
\begin{align}
\delta\mathcal{L}_{M,N,L}  =&\ \overline{\epsilon}\chi(M_{,u}+N_{,u}F+L_{,u}G)+\overline{\epsilon}\psi(M_{,v}+N_{,v}F+L_{,v}G)\,+
\\
 & -\partial_{\mu}N(\overline{\epsilon}\gamma^{\mu}\psi)-\partial_{\mu}L(\overline{\epsilon}\gamma^{\mu}\chi)+\partial_{\mu}(N\overline{\epsilon}\gamma^{\mu}\psi+L\overline{\epsilon}\gamma^{\mu}\chi)\,.\nonumber \\
\delta\mathcal{L}_{\overline{\psi}\psi}  =&-2(\overline{\epsilon}\gamma^{\mu}\psi)\partial_{\mu}vX+2(\overline{\epsilon}\psi)FX+(\overline{\psi}\psi)(\overline{\epsilon}\chi)X_{,u}
\\
\delta\mathcal{L}_{\overline{\chi}\chi}  =&-2(\overline{\epsilon}\gamma^{\mu}\chi)\partial_{\mu}uY+2(\overline{\epsilon}\chi)GY+(\overline{\chi}\chi)(\overline{\epsilon}\psi)Y_{,v}\,,
\\
\delta\mathcal{L}_{\overline{\psi}\chi}  =&-(\overline{\epsilon}\gamma^{\mu}\chi)\partial_{\mu}vZ+(\overline{\epsilon}\chi)FZ+(\overline{\psi}\gamma^{\mu}\epsilon)\partial_{\mu}uZ+(\overline{\psi}\epsilon)GZ\,-
\\
 & -\frac{1}{2}(\overline{\psi}\epsilon)(\overline{\chi}\chi)Z_{,u}-\frac{1}{2}(\overline{\psi}\psi)(\overline{\epsilon}\chi)Z_{,v}\,.\nonumber 
\end{align}
there is a single boundary contribution arising from the first term. A key simplification is the following: the terms $\overline{\chi}\chi\overline{\epsilon}\chi=0=\overline{\psi}\psi\overline{\epsilon}\psi$, since each Majorana spinor has two independent components, thus any cubic term in spinors
is zero in two spacetime dimensions. We also use a Fierz rearrangement and the fact that $\overline{\chi}\gamma^{\mu}\chi=0=\overline{\chi}\gamma^{\mu\nu}\chi$ for any $\chi$ since $t_{1}=t_{2}=-1$. Collecting the different spinorial combinations, we get for the variation of the interacting action the generic form
%
%
\small
\begin{align}
\delta\mathcal{L}_{\sf{int}} =&~~~
(\overline{\psi}\epsilon)\left(M_{,v}+(Z+L_{,v})G+(2X+N_{,v})F\right)+(\overline{\epsilon}\chi)\left(M_{,u}+(2Y+L_{,u})G+(N_{,u}+Z)F\right)-
\nonumber \\ & 
-(\overline{\epsilon}\gamma^{\mu}\chi)\left((Z+L_{,v})\partial_{\mu}v+(2Y+L_{,u})\partial_{\mu}u\right)
+(\overline{\psi}\gamma^{\mu}\epsilon)\left((Z+N_{,u})\partial_{\mu}u+(2X+N_{,v})\partial_{\mu}v\right)+
\nonumber \\ & 
+(\overline{\psi}\epsilon)(\overline{\chi}\chi)\left(-\frac{1}{2}Z_{,u}+Y_{,v}\right)+(\overline{\psi}\psi)(\overline{\epsilon}\chi)\left(-\frac{1}{2}Z_{,v}+X_{,u}\right)\,.\nonumber 
\end{align}
\normalsize
The above equation must be equal to zero for any field configuration, which implies a system of differential equations for the coupling functions $M,N,L,X,Y,Z$. Its solution can be expressed in terms of a single function
$W(u,v)$, that we will call the \emph{superpotential}, in the form
\begin{align}
&M =0\,,
&\qquad&
X =\frac{1}{2}\frac{\partial^{2}W}{\partial v^{2}}\,,
\\&
N=-\frac{\partial W}{\partial v}\,,
&\qquad&
Y=\frac{1}{2}\frac{\partial^{2}W}{\partial u^{2}}\,,
\\&L=-\frac{\partial W}{\partial u}\,,
&\qquad& 
Z=\frac{\partial^{2}W}{\partial u\partial v}\,.
\end{align}
Hence the interacting part of the action reads
\begin{align}
I_{\sf{int}} & =\kappa\int\dd^{2}x\left(-\frac{\partial W}{\partial v}F-\frac{\partial W}{\partial u}G+\frac{1}{2}\overline{\psi}\psi\frac{\partial^{2}W}{\partial v^{2}}+\frac{1}{2}\overline{\chi}\chi\frac{\partial^{2}W}{\partial u^{2}}+\overline{\psi}\chi\frac{\partial^{2}W}{\partial u\partial v}\right)\,.
\end{align}
Combining with $I_{\mathrm{free}}$ in \eqref{free} we can eliminate the auxiliary fields $F,G$ by means of their equations of motion
\begin{align}
    F=\frac{\partial W}{\partial v}
    \qquad\qquad\qquad
    G=\frac{\partial W}{\partial u}
\end{align}
obtaining for the complete action
\small
\begin{align}
    &I  =-\frac{1}{2}\int\dd^{2}x\left( 
    (\partial v)^{2}
    +
    (\partial u)^{2}
    +F^{2}
    +G^{2}
    +\overline{\psi}\not{\partial}\psi
    +\overline{\chi}\not{\partial}\chi
    -\overline{\psi}\psi\frac{\partial^{2}W}{\partial v^{2}}
    -\overline{\chi}\chi\frac{\partial^{2}W}{\partial u^{2}}
    -2\overline{\psi}\chi\frac{\partial^{2}W}{\partial u\partial v}\right)
\label{full}
\end{align}
\normalsize

\newpage

\paragraph{Supersymmetry algebra: }
In what follows we adopt the collective shorthand notation
$v_{I}  =(v,u)$, $\psi_{I}=(\psi,\chi)$ and $F_{I}=(F,G)$, which implies for the supersymmetry variations
\begin{align}
\delta v_{I}  =\overline{\epsilon}\psi_{I}\,,
\hspace{2cm} \delta\psi_{I}  =(\not{\partial}v_{I}+F_{I})\epsilon\,,  \hspace{2cm}
\delta F_{I} =\overline{\epsilon}\not{\partial}\psi_{I}\,.\qquad
\label{culoculo}
\end{align}
This allows us to write the action principle as $I=I_{\sf{free}}+I_{\sf{int}}$
where 
\begin{align}
I_{\sf{free}} & =\int\dd^{2}x\sum_{I}\left(-\frac{1}{2}(\partial v_{I})^{2}-\frac{1}{2}\overline{\psi}_{I}\not{\partial}\psi_{I}-\frac{1}{2}F_{I}^{2}\right)\,,\\
I_{\sf{int}} & =\int\dd^{2}x\sum_{I}\left(-\frac{\partial W}{\partial v_{I}}F_{I}+\frac{1}{2}\sum_{J}\overline{\psi}_{I}\psi_{J}\frac{\partial^{2}W}{\partial v_{I}\partial v_{J}}\right)
\end{align}
Preserving the boundary terms, the variation of the full action under supersymmetry transformation is given by
\begin{align}
\delta I & =\overline{\epsilon}\int\dd x\sum_{I}\partial_{\mu}\left(\frac{1}{2}\not{\partial}v_{I}\gamma^{\mu}\psi_{I}-\partial^{\mu}v_{I}\psi_{I}+\frac{1}{2}\gamma^{\mu}F_{I}\psi_{I}-\frac{\partial W}{\partial v_{I}}\gamma^{\mu}\psi_{I}\right)\,,
\end{align}
and consequently the Noether (super)current reads
\begin{align}
S^{\mu} & \equiv\sum_{I}\overline{\epsilon}\left(-\not{\partial}v_{I}+\frac{\partial W}{\partial v_{I}}\right)\gamma^{\mu}\psi_{I}\,.
\end{align}
%
This current is conserved on-shell $\partial_{\mu}S^{\mu}=0$. The supercharge is defined as usual as the space integral of the time component of the corresponding current, namely
\begin{align}
Q & \equiv\int\dd xS^{0}=\int\dd x\sum_{I}\left(-\not{\partial}v_{I}+\frac{\partial W}{\partial v_{I}}\right)\gamma^{0}\psi_{I}\,.
\end{align}
One can check that the supersymmetry variations (\ref{culoculo}) can be recovered by computing the equal time (graded) Dirac bracket between the charges $Q$ and the fields, as in $\delta v_{I}=\{v_{I}(x),\overline{\epsilon}Q\}_{D}$ and $\delta\psi_{I}=\{\psi_{I}(x),\overline{\epsilon}Q\}_{D}$. To do that, we make use of the following non-vanishing Dirac brackets
\begin{align}
\{\psi_{I\alpha}(x),\overline{\psi}_{I}{}^{\beta}(y)\}_{D} & =\delta_{IJ}(\gamma^{0})_{\alpha}{}^{\beta}\delta(x-y)\,,\label{Dirac bracket 1}\\
\{u_{I}(x),\dot{u}_{J}(y)\}_{D} & =\delta_{IJ}\delta(x-y)\,.\label{Dirac bracket 2}
\end{align}
In order to keep track of the Grassmann variables we compute the Dirac bracket between $\{\overline{\epsilon}_{1}Q,\overline{\epsilon}_{2}Q\}$ where $\overline{\epsilon}_{1}$ and $\overline{\epsilon}_{2}$ are arbitrary spinorial parameters. Direct computation leads to
\begin{align}
\{\overline{\epsilon}_{1}Q,\overline{\epsilon}_{2}Q\}_{D} & =2\,\overline{\epsilon}_{1}\gamma^{\mu}\epsilon_{2}\,P_{\mu}+\overline{\epsilon}_{1}\gamma^{0}\gamma^{1}\gamma\epsilon_{2}\,Z\,,
\end{align}
where the momentum vector $P_{\mu}$ takes the form
\begin{align}
P_{0} & =\frac{1}{2}\int\dd x\sum_{I}\left(\dot{v}_{I}^{2}+v_{I}^{\prime2}+\left(\frac{\partial W}{\partial v_{I}}\right)^{2}-\sum_{J}\overline{\psi}_{I}\psi_{J}\frac{\partial^{2}W}{\partial v_{I}\partial v_{J}}\right)\equiv E\,,\\
P_{1} & =\int\dd x\sum_{I}\dot{v}_{I}v_{I}^{\prime}\,,
\end{align}
and the \emph{central charge} $Z$ is  given by a boundary term with the form
\begin{align}
Z & =\int\dd x\, \frac{\partial}{\partial x}\left(2W(u_{I})+\frac{1}{2}\sum_{I}\overline{\psi}_{I}\psi_{I}\right)\,.
\end{align}

To better understand the meaning of this algebra, we can pick a basis of the spinor space $\theta_{+}=\left(\begin{smallmatrix}1\\
0
\end{smallmatrix}\right)\,,\theta_{-}=\left(\begin{smallmatrix}0\\
1
\end{smallmatrix}\right)$ and define $Q_{+}=\overline{\theta}_{+}Q\,,Q_{-}=\overline{\theta}_{-}Q$.
Then the algebra can be written as
\begin{align}
\{Q_{+},Q_{+}\}_{D} & =-i(E+P_{1})\,,\nonumber \\
\{Q_{-},Q_{-}\}_{D} & =-i(E-P_{1})\,,\\
\{Q_{+},Q_{-}\}_{D} & =-i Z\,.\nonumber 
\end{align}
In the quantum theory, the functions on phase space map to operators acting on the Hilbert space while the Dirac brackets maps to (anti)commutators $\{\cdot,\cdot\}_{D}\to\frac{1}{i}[\cdot,\cdot]_\pm$. Then, the resulting quantum algebra reads
\begin{align}
Q_{+}^{2} & =(E+P_{1})\,,\qquad Q_{-}^{2}=(E-P_{1})\,,\qquad Q_{+}Q_{-}+Q_{-}Q_{+}=\kappa Z\,.
\end{align}
following the argument in \cite{Olive-Witten:1978mh}, we have that
\begin{align}
(Q_{+}\pm Q_{-})^{2} & =(2E\pm Z)
\end{align}
but $(Q_{+}\pm Q_{-})^{2}\geq0$ as they are Hermitian operators, and then
\begin{align}
E & \geq\frac{1}{2}|Z|\,.
\end{align}
Particularizing to the bosonic sector, this implies that the energy must satisfy the bound
\begin{align}
E &\geq  |W(+\infty)-W(-\infty)|
\label{bogo}
\end{align}

\paragraph{Application to our case}
To embed our effective theory \eqref{eq:effective.action.BPS} in a supersymmetric model, we compare it with the form of the full supersymetric action \eqref{full} when we turn off the spinor fields $\psi$ and $\chi$. This results in the identification
\begin{align}
\left(\frac{\partial W}{\partial u}\right)^{2}+\left(\frac{\partial W}{\partial v}\right)^{2} & =2(\cos^{2}v+u^{2}\sin^{2}v)\,,
\end{align}
which is solved immediately by
\begin{align}
W(u,v) & =\sqrt{2}\,u\cos v\,.
\end{align}
The full supersymmetric action for our model then reads
$I=I_{\sf{eff}}+I_{\sf{fer}}$ with the bosonic part $I_{\sf{eff}}$ given by \eqref{eq:effective.action.BPS} and a fermionic contribution $I_{\sf{fer}}$ which according the \eqref{full} takes the form
\begin{align}
    I_{\sf{fer}}  =-\frac{1}{2}\int\dd^{2}x
    &\left( 
    \overline{\psi}\not{\partial}\psi
    +\overline{\chi}\not{\partial}\chi
    +\sqrt{2}\,u\cos v\,\overline{\psi}\psi
    +2
    \sqrt{2}\sin v\,\overline{\psi}\chi
    \right)\,.
\label{full-full}
\end{align}
The full action is then  invariant under the transformations
\begin{align}
&\delta u =\overline{\epsilon}\chi\,,
&
\delta\psi & 
=\left(\not{\partial}v-\sqrt{2}\,u\sin v\right)\epsilon\,,
\nonumber\\
&\delta v  =\overline{\epsilon}\psi\,,
&
\delta\chi & 
=\left(\not{\partial}u+\sqrt{2}\cos v\right)\epsilon\,.\label{trafos}
\end{align}

For a purely space dependent $u=u(r),v=v(r)$ and bosonic
$\psi=\chi=0$ configuration, the condition of vanishing  supersymmetry variations for a non trivial $\epsilon$ spinor requires
\begin{align}
(\gamma^{1}v^{\prime}-\sqrt{2}u\sin v)\epsilon & =0\,,
\qquad\qquad
(\gamma^{1}u^{\prime}+\sqrt{2}\cos v)\epsilon=0\,.
\end{align}
This reduces to our first order equations by imposing on the projection on the spinor $\gamma^{1}\epsilon=\pm\epsilon$ which leads to the first order equations \eqref{eq:equations.first.order.r.1}-\eqref{eq:equations.first.order.r.2} for the bosonic fields
\begin{align}
\partial_r v\mp\sqrt{2}\,u\sin v & =0\,,
\qquad \qquad
\partial_r u\pm\sqrt{2}\cos v=0\,.
\end{align}
This implies that the first order equations we have found result in solutions that are invariant under half of the supersymmetries of the supersymmetric extension of the theory. This is in agreement with the well known result about Bogomol\'{}nyi-Prasad-Sommerfield solutions. Moreover, equation \eqref{bogo} turns out to be precisely the  Bogomol\'{}nyi-Prasad-Sommerfield bound found in section \ref{sec:BPS.bound}.

\newpage
\subsection{Detailed calculations for the variations of the different terms on the Lagrangian}
The variation of the free Lagrangian:
\begin{align}
\delta\mathcal{L}_{\sf{free}} & =-\partial_{\mu}v\partial^{\mu}\delta v-\frac{1}{2}\delta\overline{\psi}\not{\partial}\psi-\frac{1}{2}\overline{\psi}\not{\partial}\delta\psi+F\delta F\,,\\
 & =-\partial_{\mu}v\partial^{\mu}(\overline{\epsilon}\psi)-\frac{1}{2}\overline{\epsilon}(-\not{\partial}v+F)\not{\partial}\psi-\frac{1}{2}\overline{\psi}\not{\partial}(\not{\partial}v+F)\epsilon+F\overline{\epsilon}\not{\partial}\psi\,,\nonumber \\
 & =-\partial_{\mu}v\overline{\epsilon}\partial^{\mu}\psi+\frac{1}{2}\overline{\epsilon}\not{\partial}v\not{\partial}\psi-\frac{1}{2}\overline{\epsilon}F\not{\partial}\psi-\frac{1}{2}\overline{\psi}\not{\partial}\not{\partial}v\epsilon-\frac{1}{2}\overline{\psi}\not{\partial}F\epsilon+F\overline{\epsilon}\not{\partial}\psi\,,\nonumber \\
 & =-\partial_{\mu}v\overline{\epsilon}\partial^{\mu}\psi+\frac{1}{2}\overline{\epsilon}\partial^{\mu}v\partial_{\mu}\psi-\frac{1}{2}\overline{\psi}\partial_{\mu}\partial^{\mu}v\epsilon-\frac{1}{2}\overline{\psi}\not{\partial}F\epsilon+\frac{1}{2}F\overline{\epsilon}\not{\partial}\psi\nonumber \\
 & +\frac{1}{2}\overline{\epsilon}\partial_{\mu}(\not{\partial}v\gamma^{\mu}\psi-\partial^{\mu}v\psi)\,,\nonumber \\
 & =\overline{\epsilon}\partial_{\mu}\left(\frac{1}{2}\not{\partial}v\gamma^{\mu}\psi-\partial^{\mu}v\psi+\frac{1}{2}\gamma^{\mu}\psi F\right)\nonumber 
\end{align}

Now we compute the variation of the interaction Lagrangian which is
the sum of the following terms
\begin{align}
\mathcal{L}_{M,N,L} & =M(u,v)+N(u,v)F+L(u,v)G\,,\\
\mathcal{L}_{\overline{\psi}\psi} & =\overline{\psi}\psi X(u,v)\,,\qquad\mathcal{L}_{\overline{\chi}\chi}=\overline{\chi}\chi Y(u,v)\,,\qquad\mathcal{L}_{\overline{\psi}\chi}=\overline{\psi}\chi Z(u,v)\,.\nonumber 
\end{align}
The variation of the term with no fermions is
\begin{align}
\delta\mathcal{L}_{M,N,L} & =\delta M+\delta NF+N\delta F+\delta LG+L\delta G\,,\\
 & =M_{,u}\delta u+M_{,v}\delta v+(N_{,u}\delta u+N_{,v}\delta v)F+N\overline{\epsilon}\not{\partial}\psi+(L_{,u}\delta u+L_{,v}\delta v)G+L\overline{\epsilon}\not{\partial}\chi\,,\nonumber \\
 & =(M_{,u}+N_{,u}F+L_{,u}G)\delta u+(M_{,v}+N_{,v}F+L_{,v}G)\delta v+N\overline{\epsilon}\gamma^{\mu}\partial_{\mu}\psi+L\overline{\epsilon}\gamma^{\mu}\partial_{\mu}\chi\,,\nonumber \\
 & =(M_{,u}+N_{,u}F+L_{,u}G)\overline{\epsilon}\chi+(M_{,v}+N_{,v}F+L_{,v}G)\overline{\epsilon}\psi-\partial_{\mu}N\overline{\epsilon}\gamma^{\mu}\psi-\partial_{\mu}L\overline{\epsilon}\gamma^{\mu}\chi\nonumber \\
 & +\partial_{\mu}(N\overline{\epsilon}\gamma^{\mu}\psi+L\overline{\epsilon}\gamma^{\mu}\chi)\,.\nonumber 
\end{align}
The variation of the $\overline{\psi}\psi$ term
\begin{align}
\delta\mathcal{L}_{\overline{\psi}\psi} & =\delta\overline{\psi}\psi X+\overline{\psi}\delta\psi X+\overline{\psi}\psi\delta X\,,\\
 & =\overline{\epsilon}(-\not{\partial}v+F)\psi X+\overline{\psi}(\not{\partial}v+F)\epsilon X+\overline{\psi}\psi(X_{,u}\delta u+X_{,v}\delta v)\,,\nonumber \\
 & =-2\overline{\epsilon}\not{\partial}v\psi X+2\overline{\epsilon}\psi FX+\overline{\psi}\psi X_{,u}\overline{\epsilon}\chi+\overline{\psi}\psi X_{,v}\overline{\epsilon}\psi\,,\nonumber \\
 & =-2\overline{\epsilon}\gamma^{\mu}\psi\partial_{\mu}vX+2\overline{\epsilon}\psi FX+\overline{\psi}\psi\overline{\epsilon}\chi X_{,u}\,,\nonumber 
\end{align}
Note that the term $\overline{\psi}\psi\overline{\epsilon}\psi$ vanishes
because we are dealing with Majorana fermions in $D=2$ which has
two independent Grassmann components.

The variation of the $\overline{\chi}\chi$ term is
\begin{align}
\delta\mathcal{L}_{\overline{\chi}\chi} & =\delta\overline{\chi}\chi Y+\overline{\chi}\delta\chi Y+\overline{\chi}\chi\delta Y\,,\\
 & =\overline{\epsilon}(-\not{\partial}u+G)\chi Y+\overline{\chi}(\not{\partial}v+F)\epsilon Y+\overline{\chi}\chi(Y_{,u}\delta u+Y_{,v}\delta v)\,,\nonumber \\
 & =-2\overline{\epsilon}\not{\partial}u\chi Y+2\overline{\epsilon}G\chi Y+\overline{\chi}\chi Y_{,u}\overline{\epsilon}\chi+\overline{\chi}\chi Y_{,v}\overline{\epsilon}\psi\,,\nonumber \\
 & =-2\overline{\epsilon}\gamma^{\mu}\chi\partial_{\mu}uY+2\overline{\epsilon}\chi GY+\overline{\chi}\chi Y_{,v}\overline{\epsilon}\psi\,,\nonumber 
\end{align}
again the term $\overline{\chi}\chi\overline{\epsilon}\chi$ vanishes. 

The variation of the mixed terms is
\begin{align}
\mathcal{L}_{\overline{\psi}\chi} & =\delta\overline{\psi}\chi Z+\overline{\psi}\delta\chi Z+\overline{\psi}\chi\delta Z\,,\\
 & =\overline{\epsilon}(-\not{\partial}v+F)\chi Z+\overline{\psi}(\not{\partial}u+G)\epsilon+\overline{\psi}\chi(Z_{,u}\delta u+Z_{,v}\delta v)\,,\nonumber \\
 & =-\overline{\epsilon}\not{\partial}v\chi Z+\overline{\epsilon}\chi ZF+\overline{\psi}\not{\partial}u\epsilon+\overline{\psi}\epsilon G+\overline{\psi}\chi Z_{,u}\overline{\epsilon}\chi+\overline{\psi}\chi Z_{,v}\overline{\epsilon}\psi\,,\nonumber \\
 & =-\overline{\epsilon}\gamma^{\mu}\chi\partial_{\mu}vZ+\overline{\epsilon}\chi ZF+\overline{\psi}\gamma^{\mu}\epsilon\partial_{\mu}u+\overline{\psi}\epsilon G+\overline{\psi}\chi\overline{\chi}\epsilon Z_{,u}+\overline{\psi}\chi\overline{\epsilon}\psi Z_{,v}\,,\nonumber 
\end{align}
Now we used the Majorana flip identities
\begin{align}
(\overline{\psi}\chi)(\overline{\chi}\epsilon) & =-\frac{1}{2}\sum_{A}(\overline{\psi}\Gamma_{A}\epsilon)(\overline{\chi}\Gamma^{A}\chi)=-\frac{1}{2}(\overline{\psi}\epsilon)(\overline{\chi}\chi)\,.\\
(\overline{\psi}\chi)(\overline{\epsilon}\psi) & =-\frac{1}{2}\sum_{A}(\overline{\psi}\Gamma_{A}\psi)(\overline{\epsilon}\Gamma^{A}\chi)=-\frac{1}{2}(\overline{\psi}\psi)(\overline{\epsilon}\chi)\,.
\end{align}
where we have used the fact that $\overline{\chi}\gamma^{\mu}\chi=\overline{\chi}\gamma^{\mu\nu}\chi=0$
due to the Majorana flip. Hence the final expression for the variation
of the mixed term is
\begin{align}
\mathcal{L}_{\overline{\psi}\chi} & =-(\overline{\epsilon}\gamma^{\mu}\chi)\partial_{\mu}vZ+(\overline{\epsilon}\chi)ZF+(\overline{\psi}\gamma^{\mu}\epsilon)\partial_{\mu}u+(\overline{\psi}\epsilon)G-\frac{1}{2}(\overline{\psi}\epsilon)(\overline{\chi}\chi)Z_{,u}-\frac{1}{2}(\overline{\psi}\psi)(\overline{\epsilon}\chi)Z_{,v}\,.
\end{align}

Now we proceed to compute the Noether current. For that we note that
a generic variation of the full lagrangian
\begin{align}
I & =\int\dd^{2}x\sum_{I}\left(-\frac{1}{2}(\partial v_{I})^{2}-\frac{1}{2}\overline{\psi}_{I}\not{\partial}\psi_{I}-\frac{1}{2}F_{I}^{2}-\frac{\partial W}{\partial v_{I}}F_{I}+\frac{1}{2}\sum_{J}\overline{\psi}_{I}\psi_{J}\frac{\partial^{2}W}{\partial v_{I}\partial v_{J}}\right)\,.
\end{align}
is given by
\begin{align}
\delta\mathcal{L} & =\sum_{I}\partial_{\mu}(-\partial^{\mu}v_{I}\delta v_{I}-\frac{1}{2}\overline{\psi}_{I}\gamma^{\mu}\delta\psi_{I})+\mathrm{eom}=\partial_{\mu}\Theta^{\mu}+\mathrm{eom}\,.
\end{align}
The boundary term evaluated in the supersymmetry transformations
\begin{align}
\delta v_{I} & =\overline{\epsilon}\psi_{I}\,,\label{aux var vI}\\
\delta\psi_{I} & =(\not{\partial}v_{I}+F_{I})\overline{\epsilon}\,,\label{aux var psiI}
\end{align}
is given by
\begin{align}
\Theta^{\mu} & =\sum_{I}\left(-\partial^{\mu}v_{I}\overline{\epsilon}\psi_{I}-\frac{1}{2}\overline{\psi}_{I}\gamma^{\mu}(\not{\partial}v_{I}+F_{I})\overline{\epsilon}\right)\,,\\
 & =\sum_{I}\left(-\partial^{\mu}v_{I}\overline{\epsilon}\psi_{I}-\frac{1}{2}\overline{\psi}_{I}\gamma^{\mu}\gamma^{\nu}\partial_{\nu}v_{I}\overline{\epsilon}-\frac{1}{2}\overline{\psi}_{I}\gamma^{\mu}\overline{\epsilon}F_{I}\right)\,,\nonumber \\
 & =\sum_{I}\overline{\epsilon}\left(-\partial^{\mu}v_{I}\psi_{I}-\frac{1}{2}\not{\partial}v_{I}\gamma^{\mu}\psi_{I}+\frac{1}{2}\gamma^{\mu}\psi_{I}F_{I}\right)\,,
\end{align}
in the last equality we used the identity $\overline{\psi}\gamma^{\mu}\gamma^{\nu}\epsilon=\overline{\epsilon}\gamma^{\nu}\gamma^{\mu}\overline{\psi}$.
On the other hand, the variation of the Lagragian under susy variation
leads to the following boundary term
\begin{align}
\delta\mathcal{L} & =\sum_{I}\overline{\epsilon}\partial_{\mu}\left(\frac{1}{2}\not{\partial}v_{I}\gamma^{\mu}\psi_{I}-\partial^{\mu}v_{I}\psi_{I}+\frac{1}{2}\gamma^{\mu}\psi_{I}F_{I}-\frac{\partial W}{\partial v_{I}}\gamma^{\mu}\psi_{I}\right)\equiv\partial_{\mu}K^{\mu}\,,
\end{align}
Hence the Noether current is defined as
\begin{align}
J^{\mu} & =\Theta^{\mu}-K^{\mu}=\sum_{I}\overline{\epsilon}\left(-\not{\partial}v_{I}\gamma^{\mu}\psi_{I}+\frac{\partial W}{\partial v_{I}}\gamma^{\mu}\psi_{I}\right)\,.
\end{align}
The supercharge can be written as
\begin{align}
Q & =\int\dd x\sum_{I}\left(\dot{v}_{I}(x)+\gamma^{0}\gamma^{1}v_{I}^{\prime}(x)+W_{,I}\gamma^{0}\right)\psi_{I}(x)\,
\end{align}
where we denote $W_{,I}=\partial W/\partial v_{I}$. Now we show that
we can recover the infinitesimal supersymmetry transformation by computing
the Dirac bracket between the field and the charges by considering
the fundamental Dirac bracket (\ref{Dirac bracket 1})-(\ref{Dirac bracket 2})
\begin{align}
\delta v_{J}(x) & =\{v_{J}(x),\overline{\epsilon}Q\}_{D}\,,\\
 & =\int\dd y\{v_{J}(x),\overline{\epsilon}\sum_{I}\left(\dot{v}_{I}(y)+\gamma^{0}\gamma^{1}v_{I}^{\prime}(y)+W_{,I}(y)\gamma^{0}\right)\psi_{I}(y)\,\}_{D}\,,\nonumber \\
 & =\int\dd y\sum_{I}\{v_{J}(x),\dot{v}_{I}(y)\}_{D}\overline{\epsilon}\psi_{I}(y)=\int\dd y\sum_{I}\delta_{IJ}\delta(x-y)\overline{\epsilon}\psi_{I}(y)\,,\nonumber \\
 & =\overline{\epsilon}\psi_{J}(x)\,.\nonumber 
\end{align}
which is (\ref{aux var vI}) and
\begin{align}
\delta\psi_{J}(x) & =\int\dd y\sum_{I}\{\psi_{J}(x),\overline{\epsilon}\left(\dot{v}_{I}(y)+\gamma^{0}\gamma^{1}v_{I}^{\prime}(y)+W_{,I}(y)\gamma^{0}\right)\psi_{I}(y)\,\}_{D}\,,\\
 & =\int\dd y\sum_{I}\{\psi_{J}(x),\overline{\epsilon}\psi_{I}(y)\dot{v}_{I}(y)+\overline{\epsilon}\gamma^{0}\gamma^{1}\psi_{I}(y)v_{I}^{\prime}(y)+W_{,I}(y)\overline{\epsilon}\gamma^{0}\psi_{I}(y)\}_{D}\,,\nonumber \\
 & =\int\dd y\sum_{I}(\{\psi_{J}(x),\overline{\psi}_{I}(y)\epsilon\}_{D}\dot{v}_{I}(y)+\{\psi_{J}(x),\overline{\psi}_{I}(y)\gamma^{1}\gamma^{0}\epsilon\}_{D}v_{I}^{\prime}(y)\,,\nonumber \\
 & -W_{,I}(y)\{\psi_{J}(x),\overline{\psi}_{I}(y)\gamma^{0}\epsilon\}_{D})\\
 & =\int\dd y\sum_{I}\left(\gamma^{0}\epsilon\dot{v}_{I}(y)+\gamma^{0}\gamma^{1}\gamma^{0}\epsilon v_{I}^{\prime}(y)-W_{,I}(y)\gamma^{0}\gamma^{0}\epsilon\right)\delta_{IJ}\delta(x-y)\,,\nonumber \\
 & =\left(\gamma^{\mu}\partial_{\mu}v_{J}(x)+W_{,J}(x)\right)\epsilon\,,\nonumber \\
 & =\left(\gamma^{\mu}\partial_{\mu}v_{J}+F_{J}\right)\epsilon\,,\nonumber 
\end{align}
is precisely (\ref{aux var psiI}). We have used the equation for
$F_{J}$ in the last equation and we used the identity
\begin{align}
  \{\psi_{\alpha}(x),\overline{\psi}(y)M(y)\epsilon\} &= \{\psi_{\alpha}(x),\overline{\psi}^{\beta}M_{\beta}{}^{\gamma}(y)\epsilon_{\gamma}\}=\{\psi_{\alpha}(x),\overline{\psi}^{\beta}(y)\}M_{\beta}{}^{\gamma}(y)\epsilon_{\gamma} = \nonumber \\
 & = \gamma^{0}M(y)\epsilon\delta(x-y)\,.
\end{align}
for $M(y)$ being a Clifford algebra valued matrix. Now we proceed to compute the bracket between charges. To divide the
computation in pieces we consider 
\begin{align}
Q  =\sum_{I}q_{I}\,, \hspace{2cm}
q_{I}  \equiv\int\dd x(\dot{v}_{I}+\gamma^{0}\gamma^{1}v_{I}^{\prime}+W_{,I}\gamma^{0})\psi_{I}\,.
\end{align}
then
\begin{align}
\{\overline{\epsilon}_{1}Q,\overline{\epsilon}_{2}Q\} & =\sum_{I,J}\{\overline{\epsilon}_{1}q_{I},\overline{\epsilon}_{2}q_{J}\}=\sum_{I}\{\overline{\epsilon}_{1}q_{I},\overline{\epsilon}_{2}q_{I}\}+\sum_{I>J}(\{\overline{\epsilon}_{1}q_{I},\overline{\epsilon}_{2}q_{J}\}+\{\overline{\epsilon}_{1}q_{J},\overline{\epsilon}_{2}q_{I}\})\,,\nonumber \\
 & =\sum_{I}\{\overline{\epsilon}_{1}q_{I},\overline{\epsilon}_{2}q_{I}\}+\sum_{I>J}(\{\overline{\epsilon}_{1}q_{I},\overline{\epsilon}_{2}q_{J}\}-\{\overline{\epsilon}_{2}q_{I},\overline{\epsilon}_{1}q_{J}\})\,,\label{pieces of bracket charges}
\end{align}
Let us compute the first term. To simplify the notation we will suppress
the index $I$ which is equivalent to set $I=1$ and the generalize
the result later.
\begin{align}
q_{1} & =\int\dd x\left(\dot{v}+\gamma^{0}\gamma^{1}v^{\prime}+W_{,v}\gamma^{0}\right)\psi\equiv A+B+C\,,\\
A & =\int\dd x\dot{v}\psi\,,\quad B=\int\dd x\gamma^{0}\gamma^{1}v^{\prime}\psi\,,\quad C=\int\dd xW_{,v}\gamma^{0}\psi\,.\nonumber 
\end{align}
the first bracket is
\begin{align}
\{\overline{\epsilon}_{1}q_{1},\overline{\epsilon}_{2}q_{1}\} & =\{\overline{\epsilon}_{1}A,\overline{\epsilon}_{2}A\}+2\{\overline{\epsilon}_{[1}A,\overline{\epsilon}_{2]}B\}+2\{\overline{\epsilon}_{[1}A,\overline{\epsilon}_{2]}C\}+\{\overline{\epsilon}_{1}B,\overline{\epsilon}_{2}B\}+\nonumber \\
 & +2\{\overline{\epsilon}_{[1}B,\overline{\epsilon}_{2]}C\}+\{\overline{\epsilon}_{1}C,\overline{\epsilon}_{2}C\}\,,
\end{align}
We compute the Dirac bracket between the auxiliary functions $A,B,C$
\begin{align}
\{\overline{\epsilon}_{1}A,\overline{\epsilon}_{2}A\} & =\int\dd x\dd y\{\overline{\epsilon}_{1}\dot{v}(x)\psi(x),\overline{\epsilon}_{2}\dot{v}(y)\psi(y)\}\,,\\
 & =\int\dd x\dd y\dot{v}(x)\dot{v}(y)\{\overline{\epsilon}_{1}\psi(x),\overline{\psi}(y)\epsilon_{2}\}\,,\nonumber \\
 & =\overline{\epsilon}_{1}\gamma^{0}\epsilon_{2}\int\dd x\dot{v}(x)^{2}\,.\nonumber 
\end{align}
\begin{align}
\{\overline{\epsilon}_{1}A,\overline{\epsilon}_{2}B\} & =\int\dd x\dd y\{\overline{\epsilon}_{1}\dot{v}(x)\psi(x),\overline{\psi}(y)\gamma^{1}\gamma^{0}\epsilon_{2}v^{\prime}(y)\}\,,\\
 & =\int\dd x\dd y[\dot{v}(x)v^{\prime}(y)\{\overline{\epsilon}_{1}\psi(x),\overline{\psi}(y)\gamma^{1}\gamma^{0}\epsilon_{2}\}+\overline{\epsilon}_{1}\psi(x)\{\dot{v}(x),v^{\prime}(y)\}\overline{\psi}(y)\gamma^{1}\gamma^{0}\epsilon_{2}]\,,\nonumber \\
 & =\int\dd x\dd y[\dot{v}(x)v^{\prime}(y)\overline{\epsilon}_{1}\gamma^{0}\gamma^{1}\gamma^{0}\epsilon_{2}+\overline{\epsilon}_{1}\psi(x)\overline{\psi}^{\prime}(y)\gamma^{1}\gamma^{0}\epsilon_{2}]\delta(x-y)\,,\nonumber \\
 & =\int\dd x[\dot{v}v^{\prime}\overline{\epsilon}_{1}\gamma^{1}\epsilon_{2}+(\overline{\epsilon}_{1}\psi)(\overline{\psi}^{\prime}\gamma^{1}\gamma^{0}\epsilon_{2})]\,,\nonumber \\
 & =\int\dd x\left[\dot{v}v^{\prime}\overline{\epsilon}_{1}\gamma^{1}\epsilon_{2}+\frac{1}{4}(\overline{\epsilon}_{1}\gamma_{*}\epsilon_{2})(\overline{\psi}\psi)^{\prime}\right]\,,\nonumber 
\end{align}
where we have used the identity $(\overline{\epsilon}_{1}\psi)(\overline{\psi}^{\prime}\gamma^{1}\gamma^{0}\epsilon_{2})=-\frac{1}{4}(\overline{\epsilon}_{1}\gamma^{1}\gamma^{0}\epsilon_{2})(\overline{\psi}\psi)^{\prime}$
which is a Fierz rearrangement plus integration by parts. Then next
one is
\begin{align}
\{\overline{\epsilon}_{1}A,\overline{\epsilon}_{2}C\} & =\int\dd x\dd y\{\overline{\epsilon}_{1}\dot{v}(x)\psi(x),-W_{,v}(y)\overline{\psi}(y)\gamma^{0}\epsilon_{2}\}\,,\\
 & =-\int\dd x\dd y[\dot{v}(x)W_{,v}(y)\{\overline{\epsilon}_{1}\psi(x),\overline{\psi}(y)\gamma^{0}\epsilon_{2}\}+\{\dot{v}(x),W_{,v}(y)\}\overline{\epsilon}_{1}\psi(x)\overline{\psi}(y)\gamma^{0}\epsilon_{2}]\,,\nonumber \\
 & =-\int\dd x\dd y[\dot{v}(x)W_{,v}(y)\overline{\epsilon}_{1}\gamma^{0}\gamma^{0}\epsilon_{2}-W_{,v,v}(y)\overline{\epsilon}_{1}\psi(x)\overline{\psi}(y)\gamma^{0}\epsilon_{2}]\delta(x-y)\,,\nonumber \\
 & =-\int\dd x[-\dot{v}W_{,v}\overline{\epsilon}_{1}\epsilon_{2}-W_{,v,v}\overline{\epsilon}_{1}\psi\overline{\psi}\gamma^{0}\epsilon_{2}]\,,\nonumber \\
 & =\int\dd x\left[\dot{v}W_{,v}\overline{\epsilon}_{1}\epsilon_{2}-\frac{1}{2}W_{,v,v}(\overline{\epsilon}_{1}\gamma^{0}\epsilon_{2})(\overline{\psi}\psi)\right]\,,\nonumber 
\end{align}
again we used Fierz rearrangement. 
\begin{align}
\{\overline{\epsilon}_{1}B,\overline{\epsilon}_{2}B\} 
 &=\int\dd x\dd yv^{\prime}(x)v^{\prime}(y)\{\overline{\epsilon}_{1}\gamma^{0}\gamma^{1}\psi(x),\overline{\psi}(y)\gamma^{1}\gamma^{0}\epsilon_{2}\}  =\int\dd xv^{\prime2}\overline{\epsilon}_{1}\gamma^{0}\epsilon_{2}\,. 
\\
\{ \overline{\epsilon}_{1}B,\overline{\epsilon}_{2}C\}
 & =-\int\dd x\dd yv^{\prime}(x)W_{,v}(y)\{\overline{\epsilon}_{1}\gamma^{0}\gamma^{1}\psi(x),\overline{\psi}(y)\gamma^{0}\epsilon_{2}\}   =\int\dd xv^{\prime}W_{,v}\overline{\epsilon}_{1}\gamma^{0}\gamma^{1}\epsilon_{2}\,.\nonumber \\
\{ \overline{\epsilon}_{1}C,\overline{\epsilon}_{2}C\} 
 & =-\int\dd x\dd yW_{,v}(y)W_{,v}(x)\{\overline{\epsilon}_{1}\gamma^{0}\psi(x),\overline{\psi}(y)\gamma^{0}\epsilon_{2}\}  =\int\dd x(W_{,v})^{2}\overline{\epsilon}_{1}\gamma^{0}\epsilon_{2}\,.\nonumber 
\end{align}
Hence the first term in (\ref{pieces of bracket charges}) is 
\begin{align}
\sum_{I}\{\overline{\epsilon}_{1}q_{I},\overline{\epsilon}_{2}q_{I}\} & =\int\dd x\sum_{I}\left[\overline{\epsilon}_{1}\gamma^{0}\epsilon_{2}\left(\dot{v}_{I}(x)^{2}+v_{I}^{\prime2}+(W_{,I})^{2}-W_{,I,I}(\overline{\psi}_{I}\psi_{I})\right)\right.+\\
 & \left.+2\overline{\epsilon}_{1}\gamma^{1}\epsilon_{2}\dot{v}_{I}v_{I}^{\prime}+\overline{\epsilon}_{1}\gamma_{*}\epsilon_{2}\left(\frac{1}{2}(\overline{\psi}_{I}\psi_{I})^{\prime}+2W_{,I}v_{I}^{\prime}\right)\right]\nonumber 
\end{align}

The missing piece is the mixed term in (\ref{pieces of bracket charges}).
Since $I\neq J$ then the only non-vanishing bracket once $q_{I}$
are replaced is $\{\dot{u}_{I}(x),W_{,J}(y)\}$. Therefore,
\begin{align}
\{\overline{\epsilon}_{1}q_{I},\overline{\epsilon}_{2}q_{J}\} & =\int\dd x\dd y\{\overline{\epsilon}_{1}(\dot{v}_{I}(x)+W_{,I}(x)\gamma^{0})\psi_{I}(x),\overline{\epsilon}_{2}(\dot{v}_{J}(y)+W_{,J}(y)\gamma^{0})\psi_{J}(y)\}\,,\\
 & =\int\dd x\dd y[\{\overline{\epsilon}_{1}\dot{v}_{I}(x)\psi_{I}(x),\overline{\epsilon}_{2}W_{,J}(y)\gamma^{0}\psi_{J}(y)\}+\{\overline{\epsilon}_{1}W_{,I}(x)\gamma^{0}\psi_{I}(x),\overline{\epsilon}_{2}\dot{v}_{J}(y)\psi_{J}(y)\}]\,,\nonumber \\
 & =\int\dd x\dd y[\{\dot{v}_{I}(x),W_{,J}(y)\}\overline{\epsilon}_{1}\psi_{I}(x)\overline{\epsilon}_{2}\gamma^{0}\psi_{J}(y)+\{W_{,I}(x),\dot{v}_{J}(y)\}\overline{\epsilon}_{1}\gamma^{0}\psi_{I}(x)\overline{\epsilon}_{2}\psi_{J}(y)]\,,\nonumber \\
 & =\int\dd xW_{,I,J}[-(\overline{\epsilon}_{1}\psi_{I})(\overline{\epsilon}_{2}\gamma^{0}\psi_{J})+(\overline{\epsilon}_{1}\gamma^{0}\psi_{I})(\overline{\epsilon}_{2}\psi_{J})]\,.\nonumber 
\end{align}
Hence
\begin{align}
\{\overline{\epsilon}_{1}q_{I},\overline{\epsilon}_{2}q_{J}\}-\{\overline{\epsilon}_{2}q_{I},\overline{\epsilon}_{1}q_{J}\} & =\int\dd xW_{,I,J}[(\overline{\epsilon}_{1}\psi_{I})(\overline{\psi}_{J}\gamma^{0}\epsilon_{2})+(\overline{\epsilon}_{1}\psi_{J})(\overline{\psi}_{I}\gamma^{0}\epsilon_{2})+\label{mixed bracket}\\
 & \hspace{2.5cm} +(\overline{\epsilon}_{1}\gamma^{0}\psi_{I})(\overline{\epsilon}_{2}\psi_{J})+(\overline{\epsilon}_{1}\gamma^{0}\psi_{J})(\overline{\epsilon}_{2}\psi_{I})]\,,\nonumber \\
 & =-2\int\dd xW_{,I,J}(\overline{\epsilon}_{1}\gamma^{0}\epsilon_{2})(\overline{\psi}_{J}\psi_{I})
\end{align}
Note that the right hand side of (\ref{mixed bracket}) is symmetric
in the interchange of $\psi_{I}$ and $\psi_{J}$, therefore once
we replace the Fierz rearrangement the only term in the sum on $\Gamma_{A}\in\{1,\gamma_{\mu},\gamma_{\mu\nu}\}$
that does not cancel are the identity terms. Putting everything together
\begin{align}
\{\overline{\epsilon}_{1}Q,\overline{\epsilon}_{2}Q\} & =\int\dd x\sum_{I}\left[\overline{\epsilon}_{1}\gamma^{0}\epsilon_{2}\left(\dot{v}_{I}(x)^{2}+v_{I}^{\prime2}+(W_{,I})^{2}-\sum_{J}W_{,I,J}(\overline{\psi}_{I}\psi_{J})\right)\right.+\nonumber \\
 & \left.+2\overline{\epsilon}_{1}\gamma^{1}\epsilon_{2}\dot{v}_{I}v_{I}^{\prime}+\overline{\epsilon}_{1}\gamma_{*}\epsilon_{2}\left(\frac{1}{2}\frac{\partial}{\partial x}(\overline{\psi}_{I}\psi_{I})+2W_{,I}\frac{\partial v_{I}}{\partial x}\right)\right]\nonumber \\
 & =2\overline{\epsilon}_{1}\gamma^{0}\epsilon_{2}P_{0}+2\overline{\epsilon}_{1}\gamma^{1}\epsilon_{2}P_{1}+\overline{\epsilon}_{1}\gamma_{*}\epsilon_{2}Z\,,
\end{align}
where we have identified the energy $P_{0}$, the momentum $P_{1}$
of the configuartion and the central charge 
\begin{align}
Z & =\int\dd x\sum_{I}\left(\frac{1}{2}\frac{\partial}{\partial x}(\overline{\psi}_{I}\psi_{I})+2W_{,I}\frac{\partial v_{I}}{\partial x}\right)=\int\dd x\frac{\partial}{\partial x}\left(2W+\frac{1}{2}\sum_{I}\overline{\psi}_{I}\psi_{I}\right)\,.
\end{align}

\end{document}